\documentclass[twocolumn,a4paper]{article}

\usepackage[affil-it]{authblk}
\usepackage[cm]{fullpage}
\usepackage{color}
\usepackage[normalem]{ulem}
\usepackage[utf8]{inputenc}
\usepackage{amsmath,amssymb,amsfonts}
\usepackage{algorithmic}
\usepackage{graphicx}
\usepackage{url}
\usepackage{multirow} 
\usepackage{booktabs}
\usepackage{soul}
\usepackage[inline]{enumitem}
\usepackage[utf8]{inputenc}

\providecommand{\keywords}[1]{\textbf{Keywords ---} #1}

\definecolor{Red}{RGB}{212,28,48}
\definecolor{Green}{RGB}{81,153,74}
\definecolor{Grey}{RGB}{162,162,162}

\providecommand{\tfix}{t_\mathrm{fix}}
\providecommand{\td}{t_\mathrm{d}}
\providecommand{\tf}{t_\mathrm{f}}

\graphicspath{{./plots/}, {./figures/}}

\begin{document}
\bibliographystyle{plain}

\title{Efficiency Near the Edge: Increasing the Energy Efficiency of FFTs on GPUs for Real-time Edge Computing}

\author[1,2]{Karel~Ad\'{a}mek}
\author[1,3]{Jan~Novotn\'y}
\author[4]{Jeyarajan~Thiyagalingam}
\author[1]{Wesley~Armour\thanks{E-mail address: \texttt{wes.armour@oerc.ox.ac.uk}}}
\affil[1]{Oxford e-Research Centre, Department of Engineering Sciences, University of Oxford, 7 Keble road, Oxford, OX1 3QG, United Kingdom}
\affil[2]{Faculty of Information Technology, Czech Technical University, Th\'{a}kurova 9, 160 00, Prague, Czech Republic}
\affil[3]{Research Centre for Theoretical Physics and Astrophysics, Institute of Physics, Silesian Univeristy in Opava, Bezru\v{c}ovo n\'{a}m\v{e}st\'{i} 13, CZ-74601, Opava, Czech Republic}
\affil[4]{Rutherford Appleton Laboratory, Science and Technology Facilities Council, Harwell Campus, Didcot, OX11 0QX, UK}

\maketitle

\begin{abstract}
The Square Kilometre Array (SKA) is an international initiative for developing the world's largest radio telescope with a total collecting area of over a million square meters. The scale of the operation, combined with the remote location of the telescope, requires the use of energy-efficient computational algorithms. This, along with the extreme data rates that will be produced by the SKA and the requirement for a real-time observing capability, necessitates in-situ data processing in an edge style computing solution. More generally, energy efficiency in the modern computing landscape is becoming of paramount concern. Whether it be the power budget that can limit some of the world’s largest supercomputers, or the limited power available to the smallest Internet-of-Things devices. In this paper, we study the impact of hardware frequency scaling on the energy consumption and execution time of the Fast Fourier Transform (FFT) on NVIDIA GPUs using the cuFFT library. The FFT is used in many areas of science and it is one of the key algorithms used in radio astronomy data processing pipelines. Through the use of frequency scaling, we show that we can lower the power consumption of the NVIDIA V100 GPU when computing the FFT by up to 60\% compared to the boost clock frequency, with less than a 10\% increase in the execution time. Furthermore, using one common core clock frequency for all tested FFT lengths, we show on average a 50\% reduction in power consumption compared to the boost core clock frequency with an increase in the execution time still below 10\%. We demonstrate how these results can be used to lower the power consumption of existing data processing pipelines. These savings, when considered over years of operation, can yield significant financial savings, but can also lead to a significant reduction of greenhouse gas emissions.
\end{abstract}


\section{Introduction}
\label{sec:intro}

The Fast Fourier Transform (FFT) is one of the most fundamental and widely used numerical algorithms in scientific computing, with applications in a diverse range of areas such as astronomy, image processing, audio and radar signal processing, numerical solvers, such as partial differential solvers, and mechanical systems~\cite{Brigham:1988:FFT_Applications}. 
The FFT is also an integral part of many data processing pipelines. 
For instance, the FFT is an important part of data processing pipelines in both image- \cite{astro:Tol:2018:imagegridding,astro:offringa:2014:WSCLEAN,astro:veenboer:2017:GPU, astro:farnes:2018:ska} and time-domain \cite{astro:dimoudi:2018:FDAS, astro:adamek:2020:improvedFDAS, astro:adamek:2019:hrms, astro:levin:2017:ska} radio astronomy.

The upcoming, next-generation radio telescope, the Square Kilometer Array (SKA), will employ such complex data processing pipelines to deliver science products that will provide new and exciting insights into our Universe. 

Previous studies, for example~\cite{cornwell2010scaling}, estimate that the SKA will require an exascale size high performance computing (HPC) system to provide us with such scientific products. Where, the computational footprint of the FFT, depending on the data processing task,  may occupy \cite{SKA_compute_est} up to 47\% of the overall computational footprint measured in floating-point operations per second (or FLOPS). This makes the FFT a critical algorithm for the SKA.

Processing the data captured by the SKA posses many challenges. The SKA will produce extremely large volumes of data at unprecedented rates. Furthermore, the telescope itself must be located in a radio-quiet area due to it's extreme sensitivity. This makes the persistent storage of all data not viable and transportation of these data to a well equipped (and suitably powered) data centre impractical. Finally, some science cases such as the study of Fast Radio Bursts (FRBs), necessitate near real-time data processing. Meaning that data has to be processed close to the instrument itself. These constraints present significant challenges to software and system engineers, they demand high fractions of peak performance of the hardware, whilst maintaining the best possible energy efficiency of both software and hardware.

To address the need to minimise the power consumption of the locally installed hardware, close attention must be paid to the energy efficiency of the data processing algorithms, specifically the FFT. Given the emphasis on lower power consumption in HPC in general, the ability to compute the FFT more efficiently is of interest to many computational domains.

The near real-time processing constraint means that the execution time of the data processing algorithms must not be increased significantly. An increase in the execution time might lead to either failure to process data on time and hence a loss of scientifically important data or increased capital and operational costs as more hardware would be needed to meet the real-time requirement.

Motivated by this, we have studied the impact of dynamic frequency scaling (DFS) on the energy efficiency and execution time of the FFT on NVIDIA GPUs using the cuFFT library \cite{NVIDIA:cuFFT}.
The GPU is the fastest and most energy efficient choice of hardware for image domain radio astronomy as shown by \cite{VeenboerRomein:2019}, with FPGAs a close second.
There are other FFT libraries for GPU's, notably, the clFFT library which uses the OpenCL framework. clFFT is not a vendor supported library and was shown by \cite{HPC:Steinbach:2017:gearshifft} to be slower than cuFFT on NVIDIA GPUs thus we have not considered it for this work.

Our exhaustive study, conducted on a range of state-of-the-art GPUs shows that careful tuning of the core clock frequency can save, in the case of the V100 GPU, up to 60\% (boost core clock frequency) of the energy consumption of the FFT. This saving can have a significant impact on two fronts: financial savings in recurrent costs, and the associated reduced $CO_2$ emission. We also show that these carefully tuned frequencies can be replaced with a single frequency that is specific to each model of GPU and chosen floating-point precision, whilst still being able to save on average up to 50\% of the FFT energy consumption (for the V100 GPU and boost core clock frequency).

\noindent The main contributions of this work are:
\begin{itemize}
\item{We have performed an in-depth investigation of cuFFT library's power consumption and execution time and how it changes with core clock frequency for a wide range of problem sizes and numerical precisions (FP16, FP32 and FP64) on five NVIDIA GPUs.}
\item{We identify an optimal core clock frequency with the highest energy efficiency for all problem sizes and numerical precisions and have shown that a single mean optimal frequency per GPU model gives similar power savings regardless of problem size.}
\item{We demonstrate how these results can be used to lower the power consumption of existing data processing pipelines.}
\end{itemize}

Whilst this work has been motivated by the SKA radio telescope, the conclusions of the work are applicable to any computational task that employs cuFFT running on NVIDIA GPUs.

\section{Background}
\label{sec:background}

Power consumption in HPC is being solved on multiple levels. From construction at the level of the cluster to new energy efficient hardware. The  power consumption of specific hardware depends on \textit{execution time}, the time taken to finish a calculation, and also on the utilization of the hardware (memory, cache, computing cores). The software itself also plays an important role in power consumption. Energy can be saved through proper software design, making software stable \cite{Meneses:2012:AssEnEfofFT} and through the use of appropriate algorithms.

However, concerns regarding energy efficiency in the modern computing landscape are not solely limited to HPC. Edge computing is becoming an increasingly important research area driven by the explosion of Internet-of-Things devices. The basic premise of edge computing is to capture and process data as close to their sources as is possible by utilising light weight processors. Because edge computing aims to process data locally, it minimizes wider latency and bandwidth needs and allows for real-time feedback. It is estimated that by 2025 around 150 billion devices will be connected and creating data in real-time \cite{IDC:2025}, with the FFT playing, not only an important role in the communication between devices, but also in processing collected data. Hence optimising the energy efficiency of the FFT on edge devices is of importance from an environmental perspective. This has motivated us to include NVIDIA's Jetson Nano in our selection of hardware since it represents NVIDIA's low power edge computing solution.

The idea behind DFS, which is part of the dynamic voltage and frequency scaling (DVFS) method, is to make hardware more energy efficient under different loads by adjusting hardware performance which is achieved by changing clock frequencies to fit the application running on it. By decreasing the clock frequency of a component we decrease its performance while increasing its utilization and thus decreasing the power consumption of a given component. For example, Trefethen et. al. \cite{TREFETHEN2013444} have investigated possible energy savings when running software on CPUs with a different number of threads, compilers and CPU clock frequencies.   

Applications can be broadly separated into two classes of performance, the first is where an application or algorithm is compute-bound. This is where the performance bottleneck of the application is the compute resource. This can be the number of floating-point operations which can be performed per second (FLOPS), but also the number of instructions which can be issued per second. The second broad category is memory bandwidth bound applications, where we have enough compute resources but we cannot supply the data through the memory bus to the computing cores quickly enough. In this case the performance is then limited by the memory bandwidth. This bandwidth limitation can occur at any level in the computers memory hierarchy, for example this might be at the level of access to the GPU main memory (called \textit{device memory}), or at the level of one of the caches.

We have investigated the cuFFT library using the NVIDIA Visual Profiler (NVVP). This shows that for all investigated problem sizes GPU kernels used by the cuFFT library are device memory bandwidth bound.

\subsection{FFT algorithm}
The one-dimensional discrete Fourier transformation (DFT) of a signal $x$ is given by 
    \begin{equation}
    \label{eqa:DFT}
        X_l=\sum_{n=0}^{N-1} x_n \exp{\left[-i2\pi \frac{nl}{N}\right]}\,,
    \end{equation}
where $X_l$ is the $l$-th element of a transformed signal, $x_n$ is the $n$-th element of an input signal, and $N$ is the transformation length or the \textit{FFT length}. 

The cuFFT library \cite{NVIDIA:cuFFT} uses the Cooley-Tukey algorithm \cite{Coo-Tuk:1965:FFT} for FFT sizes that can be decomposed as multiples of powers of primes from 2 to 127 and Bluestein's algorithm \cite{Bluestein:1970:FFT} otherwise. For longer FFT lengths the cuFFT library uses multiple GPU kernels to compute the entire FFT, which can be seen by studying the cuFFT library using the NVVP. In many cases, the Fourier transform is calculated more quickly if the FFT length is increased by padding to a more optimized length as was shown by \cite{HPC:Strelak:2018:cuFFT}. 

The two-dimensional Fourier transformation is given by the formula
    \begin{equation}
    \label{eqa:DFT2D}
        X_{l,k}=\sum_{m=0}^{M-1}\sum_{n=0}^{N-1} x_{n,m} \exp{\left[-i2\pi \left(\frac{nl}{N} + \frac{mk}{M}\right)\right]}\,,
    \end{equation}
where $x_{n,m}$, $X_{l,k}$ is now an element of a matrix of size $N \times M$. The sums in this equation can be evaluated independently which allows us to decompose the two-dimensional Fourier transformation into two sets of one-dimensional Fourier transformations. This is routinely done and it is indeed what cuFFT does when calculating higher-dimensional (2D, 3D) Fourier transformations as shown by the NVVP. Thus by investigating the energy efficiency of the one-dimensional Fourier transformation we are also investigating the energy efficiency of the higher-dimensional Fourier transforms.

\subsection{GPU architecture}
The GPU design methodology is different to that of a CPU. A CPU architecture is aimed at low latency computations, but also has lower throughput. In other words, the CPU can execute a wider range of complicated algorithms quickly, for example a complicated branching code, but the number of concurrently running tasks is small. A GPU architecture has high latency but also high throughput, on a GPU one can execute thousands of simple tasks but each task takes longer to process due to the simpler schedulers that are employed. Both platforms are broadening their focus, CPUs are adding more cores and increasing their vector lengths as GPU architectures become more complex and GPU schedulers become more sophisticated.

\begin{figure}[h!]
    \centering
    \includegraphics[width=\linewidth]{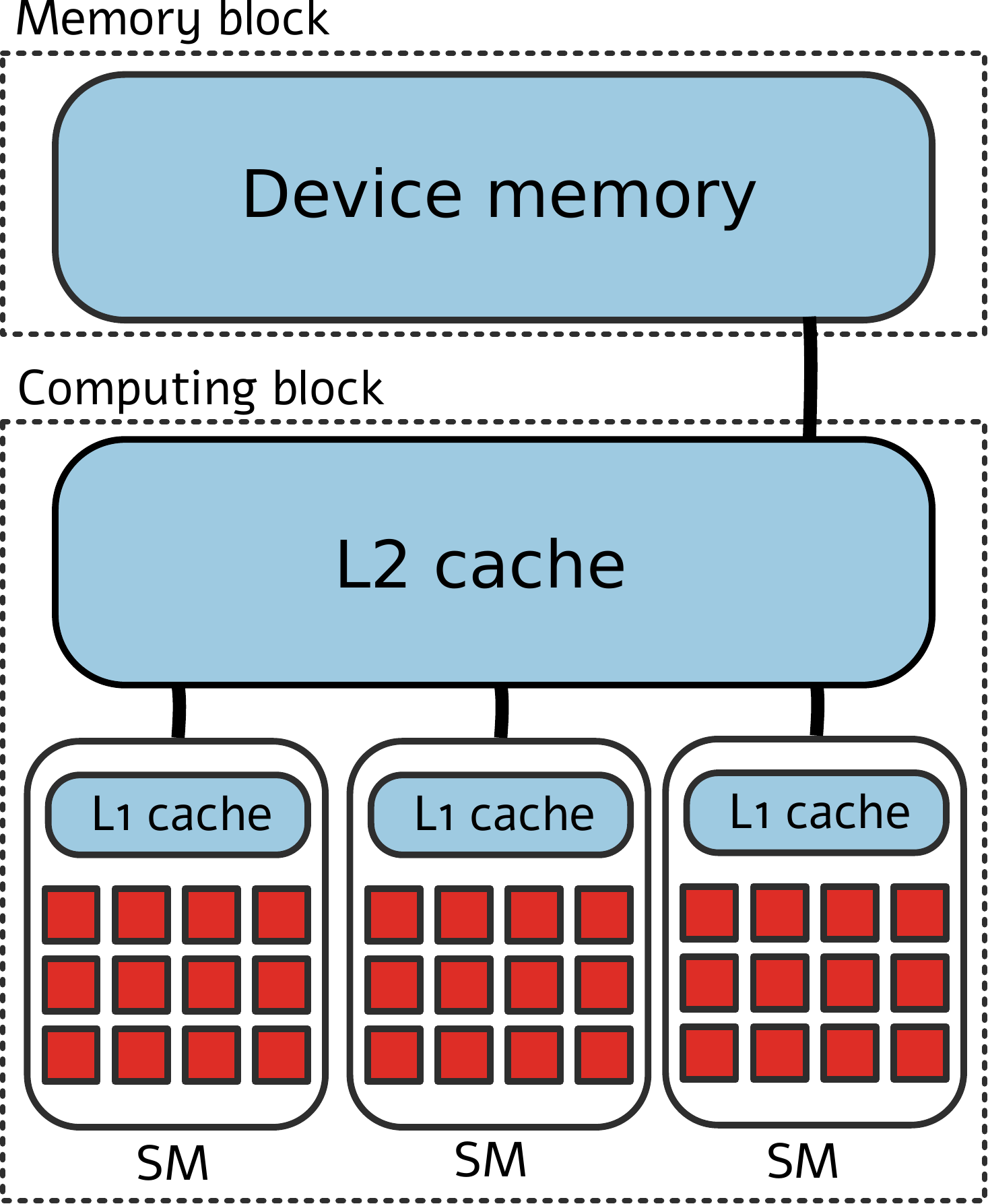}
    \caption{A schematic of the GPU architecture. \label{fig:GPUarchitecture}}
\end{figure}

The GPU architecture, which is, in simplified form, shown in Fig.~\ref{fig:GPUarchitecture}, is divided into the memory block and the compute block. The compute block is further divided into caches and streaming multiprocessors (SM) which are responsible for executing the computations. The SMs are further divided into specialized units such as floating-point cores or special function units (which are responsible for computing things like transcendental functions). The memory hierarchy on the GPU is distributed between these two blocks. The device memory which runs at the memory clock frequency has the lowest bandwidth on the GPU card and it is the memory that the CPU (host) can read/write into via the PCIe bus. The L2 cache is shared between the SMs, the L1 cache is private to each SM and the shared memory is shared amongst a group of threads called a threadblock. The L2, L1 and shared memory bandwidth is proportional to the core clock frequency, thus by using a lower core clock frequency we also decrease the bandwidth of these caches. The core clock frequency, as well as the memory clock frequency, can only be set to predefined values.

Different GPUs may use different memory modules. Amongst the tested GPUs were GPUs with GDDR memory modules (Titan XP, P4, Jetson Nano) which allow us to change the memory clock frequency, but also GPUs with HBM2 modules (Titan V, V100) which do not allow us to change the memory clock frequency.

When measuring the power consumption and performance of the GPU it is important to keep the GPU utilized. For example, the NVIDIA V100 GPU has 80 streaming multiprocessors (SM) where each SM is able to run up to 2048 threads. This gives more than 150 thousands threads which can execute concurrently. Thus in our measurements, we have used a fixed amount of data containing a different number of individual Fourier transforms to keep the GPU utilized for all tested FFT lengths.

\subsection{Real-time processing}
Data processing can be composed of a single step but more often is a series of processing steps which together form a data processing pipeline. 

The ability of the application to process data in a real-time processing scenario can be described by the real-time speed-up factor. The real-time speed-up is calculated as $S=t_a/t_p$, where $t_a$ is the time needed to acquire a given amount of data by the telescope, sensor, etc. and $t_p$ is the time taken to process that data. When $S\geq1$ the pipeline is processing data in real-time or quicker and when $S<1$ the pipeline is not managing to process data in real-time. If we assume that our toy pipeline has a real-time speed-up factor of $S=1$ that pipeline is processing the data in time but has no performance buffer to call on if needed. In such a case any increase in the execution time leads to $S<1$ and in order to process data in real-time again we must add more hardware to share the processing load. This situation is however unrealistic and a real-world pipeline would have a performance buffer to call on in the case of an unexpected event. We must also keep in mind that an increase in hardware does not necessarily equate to the same increase in a pipelines performance. The parallelization of a given task might be non-trivial, for example, communication between GPUs could be a limiting factor. In our case this approximation is appropriate as Fourier transformations which can fit into the memory of the GPU can be easily distributed amongst the GPUs.

In this work we consider two situations. The first is where the real-time processing pipeline exists and where the spare performance can be used to increase the energy efficiency of the pipeline. In the second case, we are interested in how much additional hardware is needed to process data in real-time at the best energy efficiency.

\section{Related Work}
\label{sec:related}

As of November 2019, the first two positions in the top 500 list of supercomputers are held by systems that use GPUs. Within the top ten, five systems contained GPUs. In the Green 500 list, GPUs are used in eight out of the top ten supercomputers. A clear demonstration that it is important to understand the power consumption, energy efficiency and potential energy savings for GPUs using DVFS. 

The different approaches of how to measure the power consumption, power and performance modelling and also the results of DVFS for selected applications were reviewed by Mei el al.\cite{DVFS:Mei:2016:review}. The authors note that the effect of DVFS depends not only on the architecture but also on the characteristics of the GPU application. They have found the optimal frequency for 42 GPU applications and found that 12 of them benefited from an increased core frequency compared to the default whereas for 30 applications the optimal frequency was lower than the default core frequency, and values of these optimal frequencies were different for most GPU applications. The authors called for a deeper investigation into their differences. A useful review of the DVFS technique is provided by Mittal and Vetter \cite{DVFS:Mittal:2014:review}. The review by Bridges et al. \cite{DVFS:Bridges:2016:review} looked into the modelling of the power consumption by GPUs.

A number of published studies have investigated the reliability of power measurements using internal sensors. Burtscher et al. \cite{DVFS:Burtscher:2014:K20sensor} published their experience of using built-in sensors when measuring the power consumption of NVIDIA K20 GPUs. They described several issues that they encountered when using these sensors and suggested methods to correct for these. The accuracy of the built-in sensors was investigated by Farad et al. \cite{DVFS:Farad:2019} who found that the average mean error using an abstract model of a GPU is about 10\% compared to measurements using external power meters. This error value was confirmed by Arafa et al.~\cite{DVFS:Arafa:2020:instcons} who measured the energy consumption of almost all PTX instructions for four generations of NVIDIA GPUs. They have found that the Maxwell and the Turing generations of GPUs have high energy consumption when compared to the Pascal and the Volta generations of NVIDIA GPUs which are found to be more energy efficient.

There are a number of papers where authors have used DVFS in the context of GPUs \cite{DVFS:Abe:2014, DVFS:Sethia:2014, DVFS:Wang:2020, DVFS:Loghin:2018, DVFS:Ge:2013, DVFS:Chau:2017, DVFS:Lee:2011, DVFS:GUERREIRO:2019, DVFS:Jiao:2010, DVFS:Mei:2016:review, DVFS:Tang:2019:Deep}. Guerreiro et. al.~\cite{DVFS:GUERREIRO:2019} classified GPU applications into four different categories which describe their behaviour when DVFS is applied. These categories are an extension of the compute-bound, memory-bound distinction. The early work on GPU power consumption and DVFS was performed by Jiao et al. \cite{DVFS:Jiao:2010}. They investigated the behaviour of several GPU applications which included the FFT algorithm, however, the cuFFT library was not studied because there were better performing FFT implementations at the time. The FFT was also indirectly included in Mei et al. \cite{DVFS:Mei:2016:review} as part of the convolution, and in Tang et al.  \cite{DVFS:Tang:2019:Deep} where the author investigated the effect of DVFS on deep learning applications.

In relation to radio astronomy and the SKA, there are several works. Price et al. \cite{ASTROINF:Price:2016:SKA} made a detailed investigation into power consumption, voltage and frequency scaling of the GPU implementation of the correlator for the SKA. The power consumed by the GPU in the domain of radio astronomy was investigated by Romein \cite{RADIO:Romein:2016}. The performance of the cuFFT library was investigated by Jondra et al. \cite{HPC:Jodra:2015} along with its power consumption. However, increases in energy efficiency due to frequency scaling were not investigated.

	\section{Experimental Setup and Evaluation}
\label{sec:evaluation}

The code that we have used\footnote{can be found at \url{https://github.com/KAdamek/cuFFT_benchmark}} for measurements of the energy efficiency of the FFT algorithm consists of a basic implementation of the NVIDIA cuFFT library \cite{NVIDIA:cuFFT}.  

The code first generates input data as pseudo random numbers on the host and then we transfer the data from the host to the device via the PCIe bus. The code runs the FFT algorithm on the GPU multiple times whilst the power used by the GPU is measured as described below. The measurements gained from multiple runs are used to calculate a relative standard deviation which we use to represent the measurement error in the results presented. We provided the GPU with enough data to ensure that it is fully utilized. The Fourier transform used was an out-of-place one-dimensional transform as provided by cuFFT. When the FFT algorithm ends the measurement of the power is stopped. Thus only the power consumption of the FFT algorithm on the GPU is measured. The calculated result is transferred back to the host. The result is then compared to the result from the same transformation again performed by the GPU, but this time using the GPU's default settings. This is done to ensure correctness. 

To technically achieve the above scenario we log the time\-stamp, power consumption, current core clock frequency and current memory clock frequency. For that we use the NVIDIA System Management Interface (\texttt{nvidia-smi}) for all GPU cards except the Jetson Nano, where we have used the \texttt{tegrastats} utility. For both we have specified the measurement interval to be 10\,ms as our tests have showed that a setting of time sampling below 10\,ms does not lead to an improvement in the time resolution of our data. The actual time between samples varied and the actual achieved sampling rate from the driver is on average 14.2\,ms for all tested FFT lengths and cards. This sampling rate fulfills the criterion of at least 15~ms (66.7~Hz) recommended by Burtscher et al.~\cite{DVFS:Burtscher:2014:K20sensor} to accurately measure the energy consumption of real-world kernels.

\begin{figure}[h!]
    \centering
    \includegraphics[width=\linewidth]{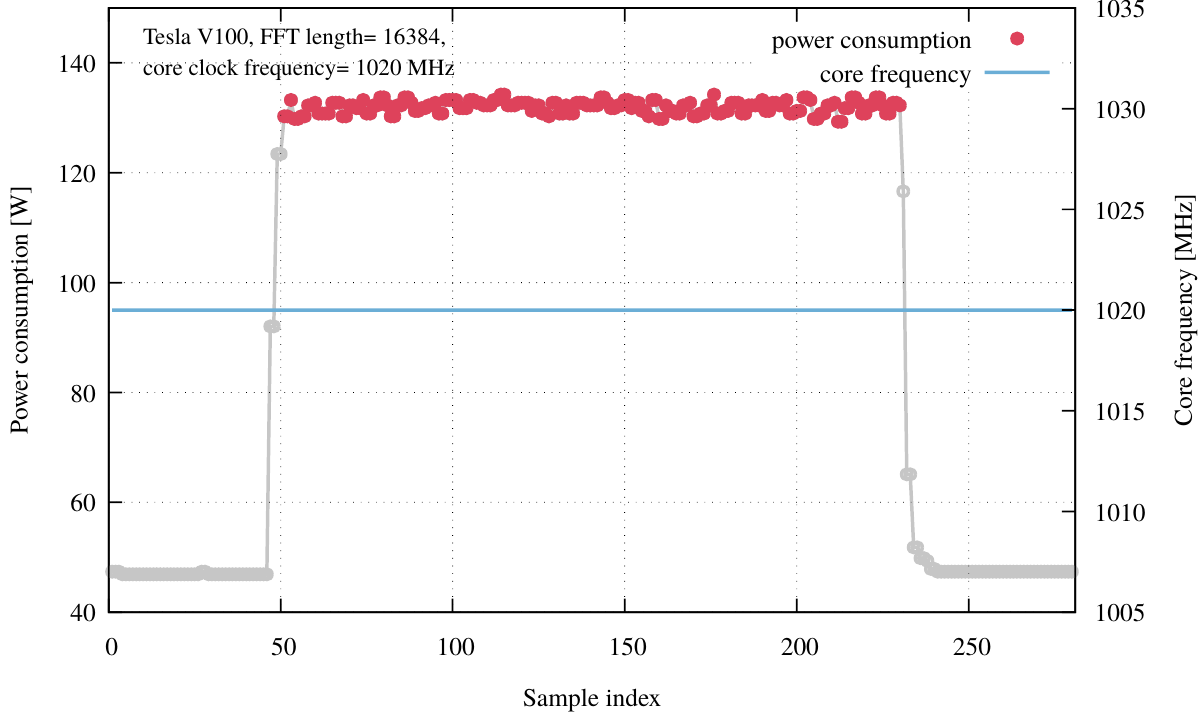}\\
    \includegraphics[width=\linewidth]{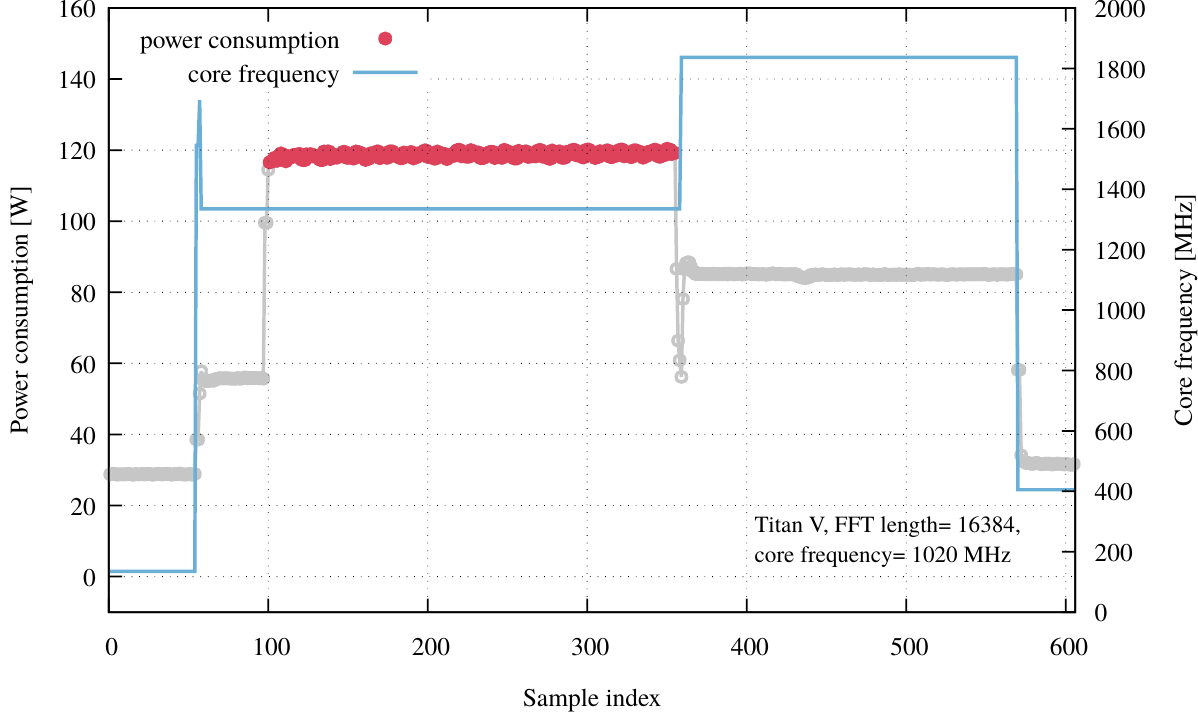}
    \caption{\label{fig:r_measurement}Parts of the log file with the GPU kernel highlighted (red dots) by the R~script between the two non-computing parts of the GPU run (grey line dots) showing the reported power consumption. The blue line corresponds to the measured core clock frequency. Specifically, the data displayed are from measurements on the Tesla V100 (top) and Titan~V (bottom) for an FFT length of $2^{14}$, single precision and the core clock frequency set to 1020~MHz (Tesla~V100) and 1912~MHz (Titan~V).}
\end{figure}

For the localization of the FFT algorithm and establishing the execution time we have used the \texttt{nvprof} utility where we have included the timestamp. Finally we log the beginning and end of each GPU kernel execution to a file. This way we produce two files containing all of the needed metrics for all possible combinations of core clock frequencies for a specific FFT length, bit precision and GPU card. The final combination (via the timestamp comparison) of these files is done by using a simple R script. Here we compute all other metrics including energy efficiency, optimal clock frequency, mean optimal core clock frequency and computational performance. In the script we also verify that the current core clock frequency is the same as the requested one, and compare the measured execution time from \texttt{nvprof} with the log timestamps of the \texttt{nvidia-smi} query. 
Using this method we have found that, for the Titan~V, the core clock frequency is capped to 1335~MHz by the driver\footnote{driver version 450.36.06} during the computation, but during the copy of the results is set to a higher core clock frequency (1837~MHz). For frequencies lower than 1335~MHz, no capping is observed. An example of the GPU kernel power consumption and active core clock frequency, which was localized using log file timestamps, is shown for the V100 GPU in Fig.~\ref{fig:r_measurement} (top). An example of the frequency capping on the Titan V GPU is shown in Fig.~\ref{fig:r_measurement} (bottom).

The choice of clock frequencies for both the memory bus and the computational cores are limited to a set of supported frequencies defined by the hardware itself. The supported core clock frequency can easily be changed by the driver API. The allowed clock frequencies of the device memory bus are limited or not changeable depending on the memory type. Since the cuFFT library is completely limited by device memory bandwidth this suggests that lowering the memory frequency would not lead to substantial increases in the energy efficiency. Thus, we have not changed the memory clock frequency in this work. Moreover the High Bandwidth Memory (HBM) which is present on the newest GPU cards (Titan~V, Tesla~V100) operates on a fixed memory clock frequency. The ranges and step sizes of the core clock frequencies that we have used are summarized in Table~\ref{tab:list_core_frequencies}.

\begin{table}[h]
\centering
\caption{\label{tab:list_core_frequencies}List of the allowed core clock frequencies from maximal $f_\mathrm{max}$ up to minimal $f_\mathrm{min}$ frequency for all cards and their corresponding frequency step size ($f_\mathrm{step}$). The size of the frequency step alternates between values shown in the column $f_\mathrm{step}$ with the exception of the Jetson Nano.}  
\resizebox{\linewidth}{!}{
\begin{tabular}{|l|l|l|l|}
\hline
Card name   & $f_\mathrm{max}$ [MHz] & $f_\mathrm{min}$ [MHz]& $f_\mathrm{step}$ [MHz] \\
\hline
Tesla V100  & 1530   & 135    & 7, 8   \\
Tesla P4    & 1531   & 455    & 12, 13 \\
Titan XP    & 1911   & 379    & 12, 13 \\
Titan V     & 1912   & 135    & 7, 8   \\
Jetson Nano & 921.6  & 76.8   & 76.8 \\
\hline
\end{tabular}}
\end{table}

The energy for a specific core clock frequency is defined as 
\begin{equation}
    E_{f} = \sum_{i} P_i \cdot t_i\,,\label{eq:energy_clock}
\end{equation}
where $P_i$ corresponds to the reported power for a sample index $i$ and $t_i$ is the time between the current sample and the previous one. Then the energy efficiency for a specific core clock frequency is given as
\begin{equation}
    E_{\mathrm{ef}} = C_{\mathrm{p}} \cdot t / E_f\,,\label{eq:efficiency}
\end{equation}
where $t$ corresponds to the time of the whole run of the computation, $E_f$ is the energy and $C_{\mathrm{p}}$ is the computational performance in FLOPS given by
\begin{equation}
    C_\mathrm{p} = \left[5 N\log_{2} (N)\cdot N_\mathrm{b} \cdot N_\mathrm{FFT}\right]/ t\,,
\end{equation}
where $N_\mathrm{b}$ is the number of FFT runs of length $N$ and $N_\mathrm{FFT}$ is the number of FFTs computed per run. The number of Fourier transforms performed ($N_\mathrm{FFT}$) depends on the FFT size as follows
\begin{equation}
N_\mathrm{FFT}=M_\mathrm{GB}/(N\cdot\mathrm{B})\,,\label{eq:n_fft}    
\end{equation}
where $M_\mathrm{GB}$ is the desired amount of memory used for FFTs in GB and $B$ is the byte size of the input data type. The optimal core clock frequency for a specific FFT length is then found as the one with the minimal consumed energy.

We define the increase in energy efficiency as 
\begin{equation}
    I_\mathrm{ef} = E_\mathrm{ef,o}/E_\mathrm{ef,d}\,,\label{eq:increase}
\end{equation}
where $E_\mathrm{ef,o}$ and $E_\mathrm{ef,d}$ are the energy efficiencies for the optimal frequency and the boost frequency respectively (given by~\eqref{eq:efficiency}).

\begin{figure}[h!]
    \centering
    \includegraphics[width=\linewidth]{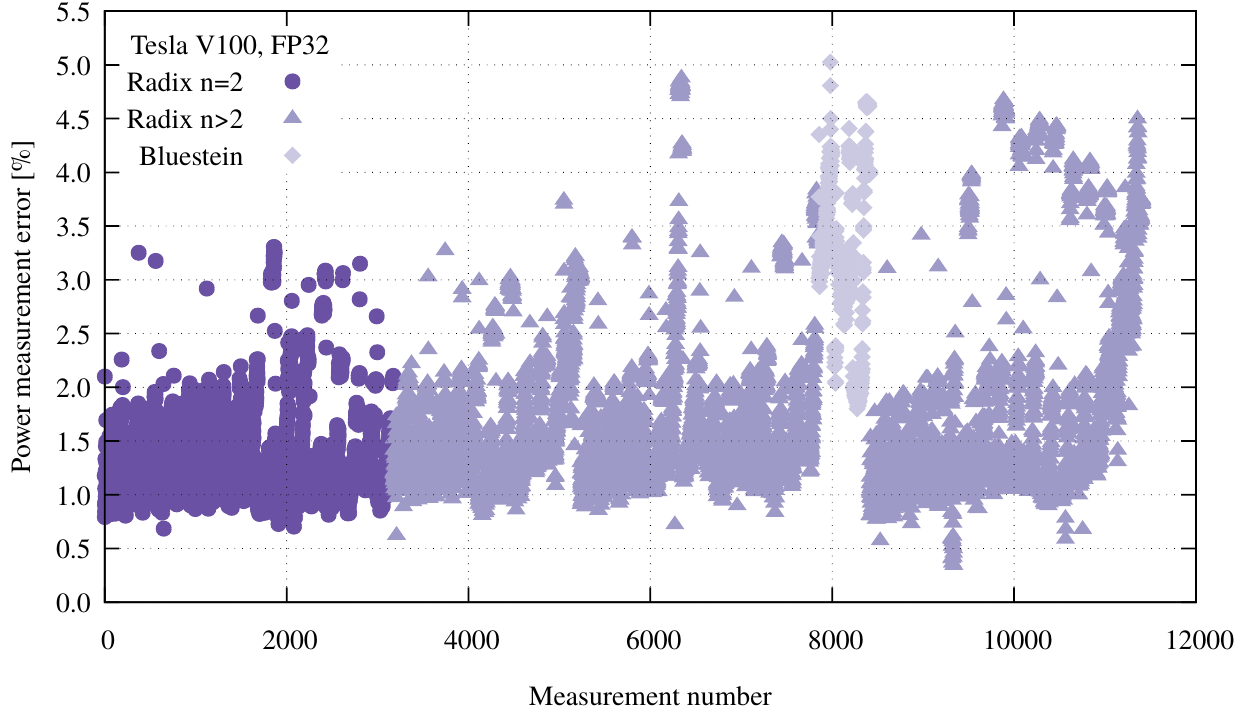}\\
    \includegraphics[width=\linewidth]{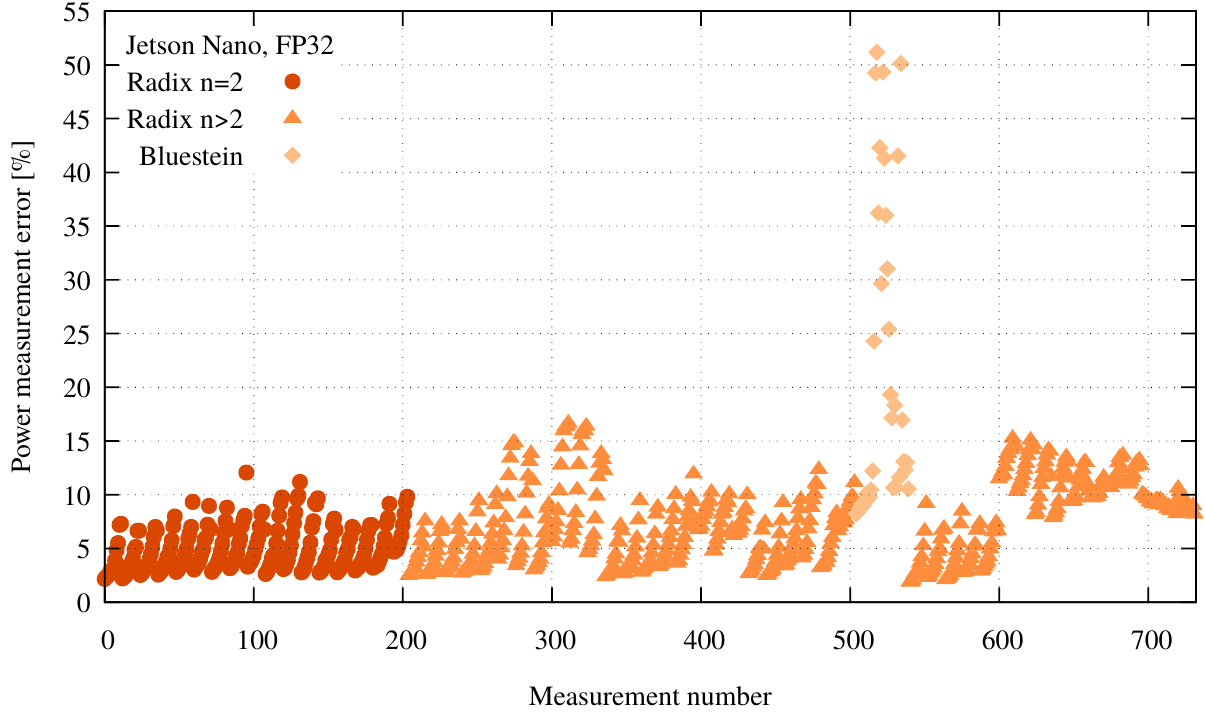}
    \caption{\label{fig:m_errors}Measurement error (V100 GPU at the top, Jetson Nano at the bottom) for all tested FFT lengths at all tested core clock frequencies.}
\end{figure}

The measurement error, that is the relative standard deviation, for the V100 GPU and the Jetson Nano is shown in Fig.~\ref{fig:m_errors}.
We have observed that the measurement error, in general, is around 5\% for all cards except the Jetson Nano. 
The GPU cards use instrumentation amplifiers for the current/voltage/power monitors, hence the potential error in the measurement is expected to be around 3--5\% \cite{noauthor_power_2018}.
The results of our power measurement correspond to the expected characteristics of the on-board chips.

For Fourier transformations of higher radices (7+) or for Fourier transformations which use the Bluestein algorithm we observe a measurement error of up to 5\%. The measurement error increases with decreasing core clock frequency and increasing number of GPU kernels used for the FFT calculation.

The measurement error for the Jetson Nano is usually below 15\% for all FFT lengths, and is below 10\% for power-of-two FFT lengths. The highest measurement error that we have observed is for Bluestein FFT lengths. For these lengths, cuFFT uses multiple kernels (for $N=139^2$ eleven GPU kernels are used) thus the high measurement error is due to the different loads these GPU kernels exert on the GPU and also the differing power consumption between them. The Bluestein FFT lengths represent a marginal case. Due to large measurement errors for Bluestein FFT lengths on the Jetson Nano we have not included these measurements into our calculations of mean optimal frequency. However, we present these results for the sake of completeness.

For the measurement of the execution time we have used the NVIDIA Visual Profiler. Using this we have found that the measurement error for the execution time was below 0.3\%.

Using propagation of uncertainty the error of the energy~\eqref{eq:energy_clock} is dominated by the measurement error of the power consumption. Based on that, the error in the increase in energy efficiency~\eqref{eq:increase} is given by

\begin{equation}
\sigma_\mathrm{R}(I_\mathrm{ef}) = \sqrt{2}\sigma_R(E_\mathrm{ef})\,,\label{eq:increase_error}
\end{equation}
where $\sigma_\mathrm{R}$ is the relative error and we have assumed that the relative error in $E_\mathrm{ef,o}$ and $E_\mathrm{ef,d}$ are equal. This gives an error for the increase in the energy efficiency of 7\% for all GPUs except the Jetson Nano where the error is 21\%. These values represent the worst case scenario since most of measurement errors are well below these values.

\section{Results}
\label{sec:results}

For our investigation, we have used five different NVIDIA GPUs from three recent architecture generations, namely V100 (Volta), Tesla~P4 (Pascal), Jetson Nano (Maxwell), Titan~V (Volta) and Titan~XP (Pascal). The relevant hardware specifications can be found in Table \ref{tab:hardware}. Both the V100 GPU, and Tesla P4 GPU are aimed at scientific applications, the P4 GPU also offers improved energy efficiency for it's generation. The Jetson Nano is a low-powered all-in-one solution for autonomous systems. The Titan~V and Titan~XP are consumer grade GPUs. 

\begin{table*}[htbp]
	\caption{GPU card specifications. The shared memory bandwidth is calculated as $\mathrm{BW (bytes/s)} = \mathrm{(bank\, bandwidth\, (bytes))} \times \mathrm{(clock\, frequency\, (Hz))} \times \mathrm{(32\, banks)} \times \mathrm{(\#\, multiprocessors)}$.}
	\begin{center}
	\resizebox{\linewidth}{!}{
	\begin{tabular}{|l|r|r|r|r|r|}
	\hline
	\textbf{} & \textbf{Titan XP} & \textbf{Tesla P4} & \textbf{Titan V} & \textbf{Tesla V100} & \textbf{Jetson Nano}\\
	\hline
	CUDA Cores & 3840 & 2560 & 5120 &  5120 & 128\\
	SMs & 30 & 20 & 80 & 80 & 2\\
	Base/Boost Core Clock & 1405/1480\,MHz & 810/1063\,MHz & 1220/1455\,MHz & 1200/1455\,MHz & 921\,MHz\\
	Memory Clock & 5005\,MHz & 3003\,MHz & 850\,MHz & 877\,MHz & 1600\,MHz\\
	Dv. m. bandwidth & 547\,GB/s & 192\,GB/s & 652\,GB/s & 900\,GB/s & 25.6\,GB/s\\
	Memory modules & GDDR5 & GDDR5 & HBM2 & HBM2 & LPDDR4\\
	Shared m. bandwidth & 5395\,GB/s & 2657\,GB/s & 14550\,GB/s & 14550\,GB/s & 230\,GB/s\\
	Memory size & 12\,GB &  8\,GB & 12\,GB & 16\,GB & 4\,GB\\
	TDP & 250\,W & 75\,W & 250\,W & 300\,W & 5/10\,W\\
	CUDA version & 10.0.130 & 10.0.130 & 10.0.130 & 10.0.130 & JetPack 4.2 SDK\\
	\hline
	\end{tabular}}
	\label{tab:hardware} 
	\end{center}
	\end{table*}

GPUs have two different frequency settings: a base and a boost core clock frequency. If not stated otherwise, we have used the boost core clock frequencies. This is because the GPU's default behaviour is to perform calculations at the boost core clock frequency. This is indeed what is observed when the GPU is set to default mode and we run our cuFFT code. When reporting energy efficiency, we use both frequencies as there is a non-linear dependency of the power consumption of a GPU on the core clock frequency. 

We have measured the complex-to-complex (C2C) one-dimensional transform for three different floating-point precisions; double (FP64), float (FP32) and half (FP16). The Tesla P4, Titan XP and Jetson Nano GPUs have limited support for the double precision format. Furthermore, the Tesla P4 and the Titan XP do not support the half (FP16) floating-point precision. In addition, when using half precision (FP16), the cuFFT library supports only power-of-two FFT lengths. 

We have investigated various FFT lengths but focused on lengths that are powers-of-two because FFT algorithms are not only best suited to processing such lengths, but also offer superior execution time performance with powers-of-two lengths. When calculating non-power-of-two FFT lengths it is often faster~\cite{HPC:Strelak:2018:cuFFT} to pad the data which needs to be Fourier transformed to the nearest higher power-of-two FFT length and then Fourier transform.

First, we present execution times for processing a fixed amount of data $\tfix$ which offers an insight into the level of optimization provided by the cuFFT library. The memory requirements to store the data needed for the Fourier transform grows linearly with the FFT length $N$. Since the cuFFT library is limited by the device memory bandwidth, the execution time consists of the time required to transfer the data to computing cores and to store the result back to the device memory $t_i$, and the time required for any additional overhead accesses to the device memory $t_o$. If the performance limiting factor is different to the device memory bandwidth, we are unable to make such a distinction in this work. In an ideal case where we would have a large enough cache, the execution time of the Fourier transform would be equal to the time $t_i$. However, because the cache size is limited, the time $t_o$ will be non-zero and directly indicate the efficiency of the implementation. By fixing the amount of memory being processed, the time $t_i$ will be a constant and any increase in the execution time of the Fourier transform will be due to time $t_o$. 

\begin{figure}[h!]
    \centering
    \includegraphics[width=\linewidth]{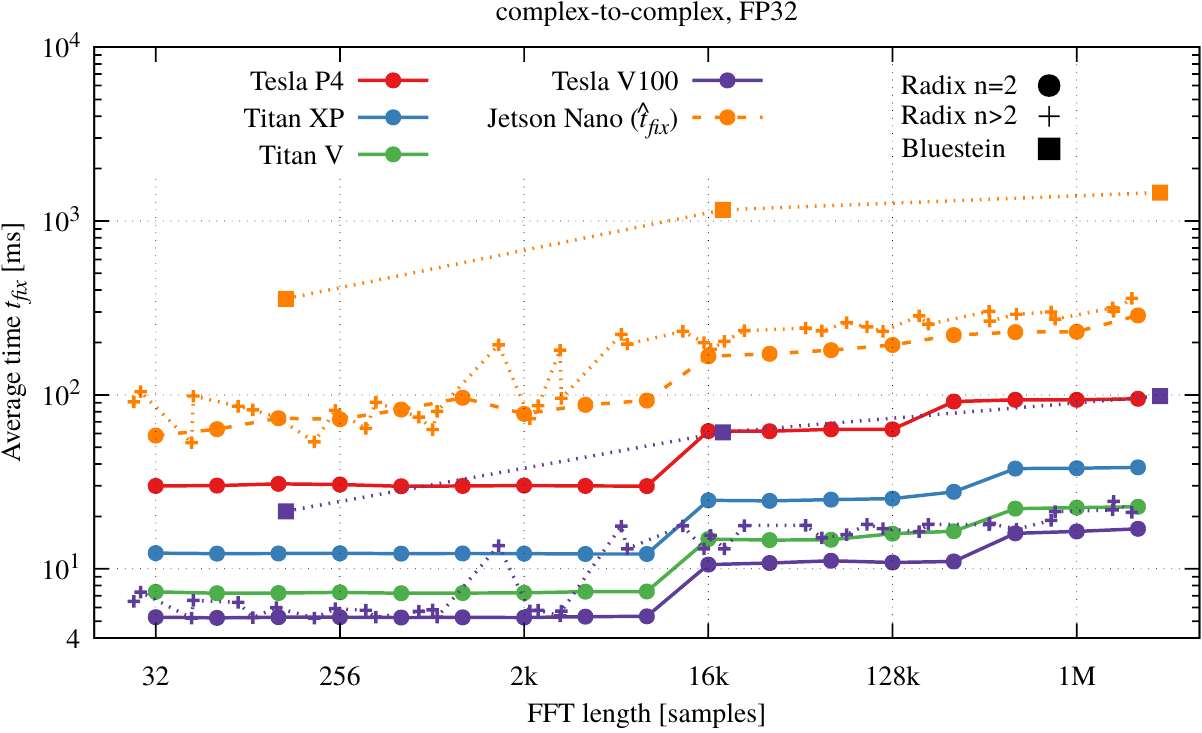}
    \caption{The execution time $\tfix$ (for FP32) required to process a fixed amount of data for different FFT lengths. The discontinuities in the execution time indicate a change of optimised GPU kernel that is used to calculate the FFT. Results for the Jetson Nano are for one quarter of the memory size.\label{fig:fixedmemExTFP32}}
\end{figure}

\begin{figure}[h!]
    \centering
    \includegraphics[width=\linewidth]{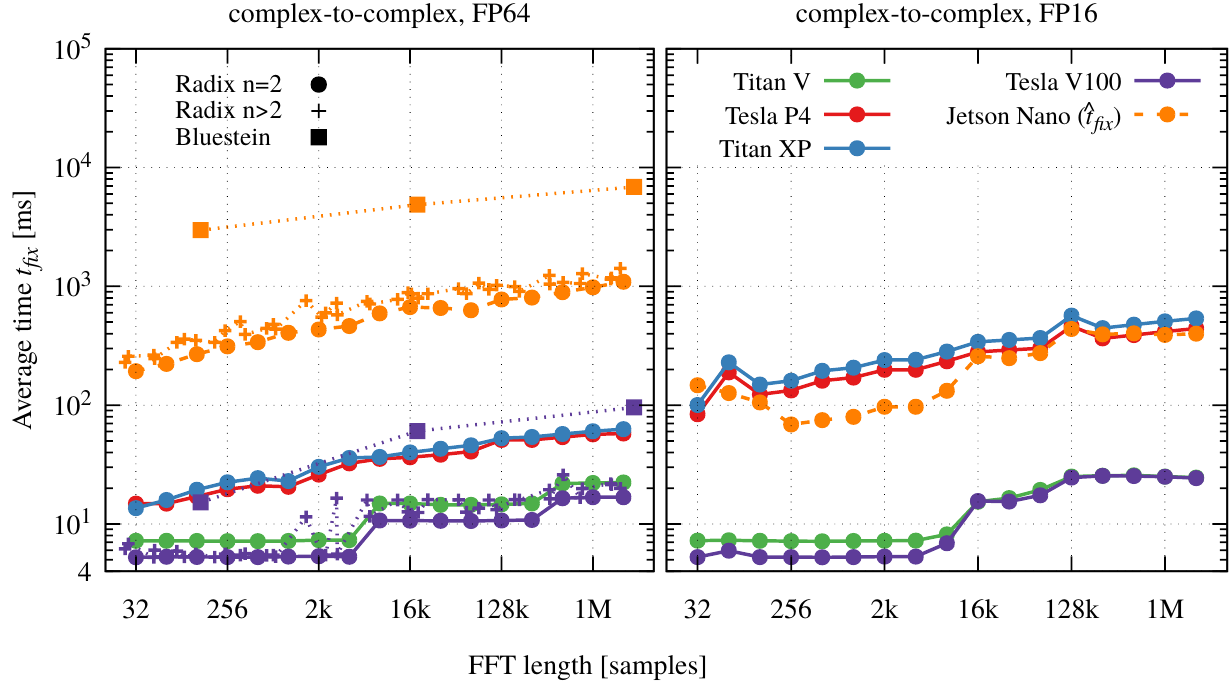}
    \caption{The execution time $\tfix$ (for FP16 and FP64) required to process a fixed amount of data for different FFT lengths. The discontinuities in execution time indicate a change of optimised GPU kernel that is used to calculate the FFT. Results for the Jetson Nano are for one quarter of the memory size.  \label{fig:fixedmemExTFP16FP64}}
\end{figure}

If we fix the amount of data that is processed then the number of FFTs performed $N_\mathrm{FFT}$ depends on the FFT length as given by \eqref{eq:n_fft}. The execution time of a single FFT within a batch is given as $t_t=\tfix/N_\mathrm{FFT}$. The execution time $\tfix$ for processing a fixed amount of data for various FFT lengths is shown in Fig.~\ref{fig:fixedmemExTFP32} for FP32 and in Fig.~\ref{fig:fixedmemExTFP16FP64} for FP16 and FP64 precision. The execution time for the Jetson Nano is for $1/4$ of the amount of data so the comparable value of $\tfix$ is $\tfix = 4\hat{t}_\mathrm{fix}$. This is due to the low amount of available memory on the card.

The execution time $\tfix$ increases in proportion to the length of the Fourier transform. However, we see regions of the same execution time with sudden increases after specific FFT lengths. These abrupt changes represent a transition from one optimized GPU kernel to another as is shown by the NVIDIA profiler. We must take these changes into account in our analysis since these GPU kernels might behave differently. When the execution time $\tfix$ does not increase for a given range of problem sizes (for example from FFT length $N=32$ to $N=8192$) it means that the higher number of floating-point operations which come with a larger problem size utilizes GPU resources other than the device memory bandwidth. Given that the Titan XP, Tesla P4 and Jetson GPUs do not fully support all tested floating-point precisions the execution time of Fourier transformations on these GPUs exhibit different behaviours.

In this work, results are presented per \textit{FFT batch}, which is the number of FFT's of a given length which fit into the fixed amount of memory that we have chosen to work with. However, most of our results, such as energy efficiency, are independent of the number of FFTs calculated provided that the GPU is fully utilised. The execution time for different core clock frequencies is denoted by $\tf$. The execution time with boost frequency is denoted as $\td$ and is taken as the execution time for the default settings. Furthermore, we have focused our discussion on the V100 GPU as it is the most current (and widely used) scientific GPU and the Jetson Nano as it represents NVIDIA's low power edge computing solution. We point out any deviations from these behaviours in the other tested GPUs when they occur.


\subsection{Frequency Scaling}
\begin{figure}[h!]
    \centering
    \includegraphics[width=\linewidth]{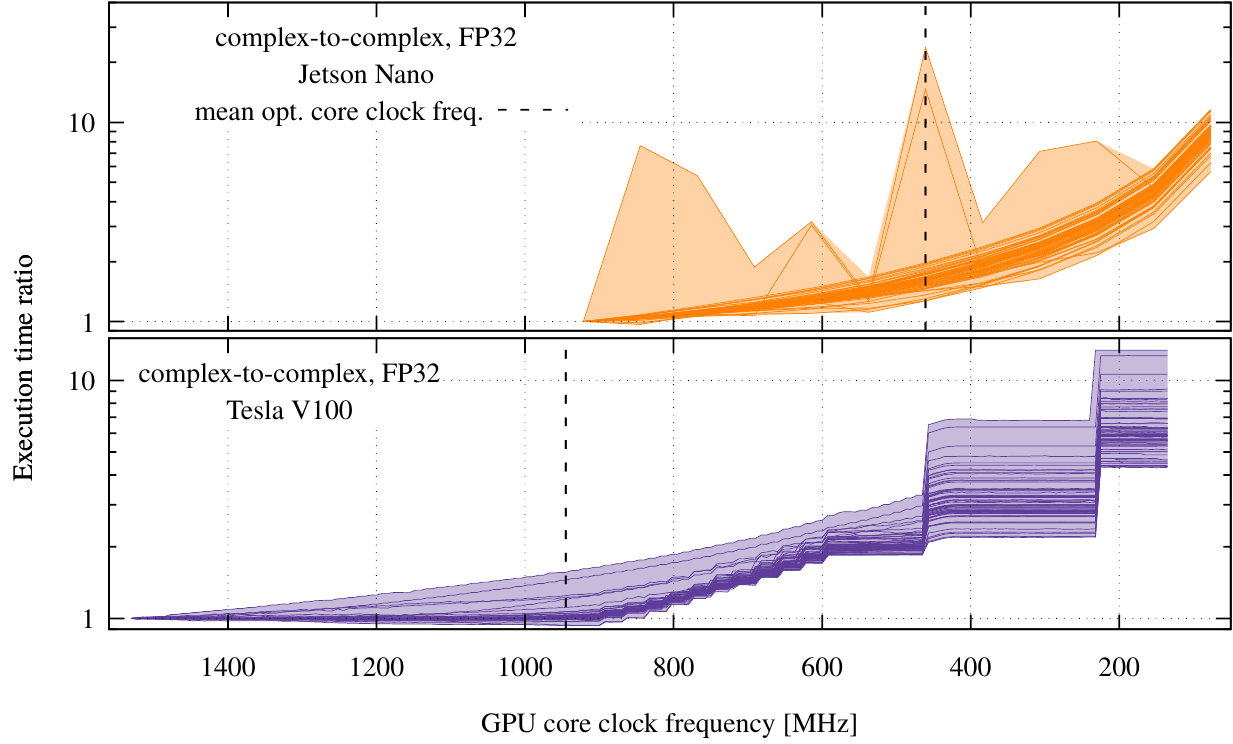}
    \caption{Ratio of the execution time $\tf$ over the default execution time $\td$ measured for the V100 GPU and the Jetson Nano. Every investigated FFT length is shown and represented by a single line. \label{fig:timefreqdep}}
\end{figure}

First, we present the behaviour of the execution time with changing core clock frequency. This is shown as a ratio of execution time $\tf$ over default execution time $\td$ in Fig.~\ref{fig:timefreqdep}, which shows all tested configurations for FP32 precision.

There are three distinct behaviours, the execution time is: 
\begin{enumerate}[label={\alph*)}]
\item decreasing at first;
\item slightly increasing;
\item increasing notably with each frequency decrease.
\end{enumerate}
In the case of the V100 GPU, the first two behaviours a) and b) are in the majority. For a few specific FFT lengths (notably for $N=8192$) we have observed behaviour c). We have observed this behaviour throughout multiple measurements and always for the same FFT lengths. Other tested GPUs behaved similarly to the V100 GPU.

The Jetson Nano exhibits a different behaviour, where most of the configurations belong to case c) with notable peaks which are present for Bluestein FFT lengths.

\begin{figure}[h!]
    \centering
    \includegraphics[width=\linewidth]{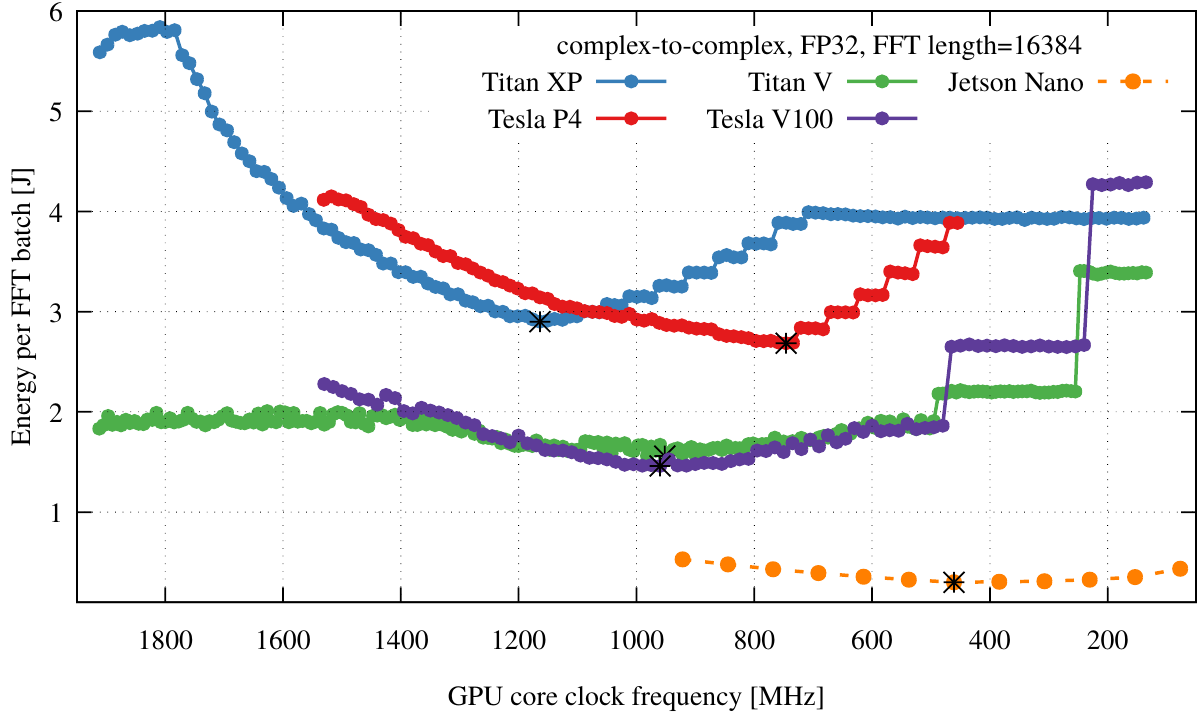}
    \caption{The energy consumed per FFT batch changes with core clock frequency. The minimum, emphasized by a black star for each tested GPU, represents the most efficient configuration and the value of the optimal frequency.  \label{fig:energyfreqdep}}
\end{figure}

The energy consumed per FFT batch calculated by equation~\eqref{eq:energy_clock} with fixed length $N=16384$ for different GPUs is shown in Fig.~\ref{fig:energyfreqdep}. For the measurement, we have used a batch of 16384 FFTs (in the case of FP32 this represents 2~GB of input data) in order to fully saturate the GPU. Notably, the energy per FFT batch on the Titan V GPU does not change above 1335~MHz. This is because the card does not run at the user selected frequency but is capped by the driver to 1335~MHz. 

\begin{figure}[h!]
    \centering
    \includegraphics[width=\linewidth]{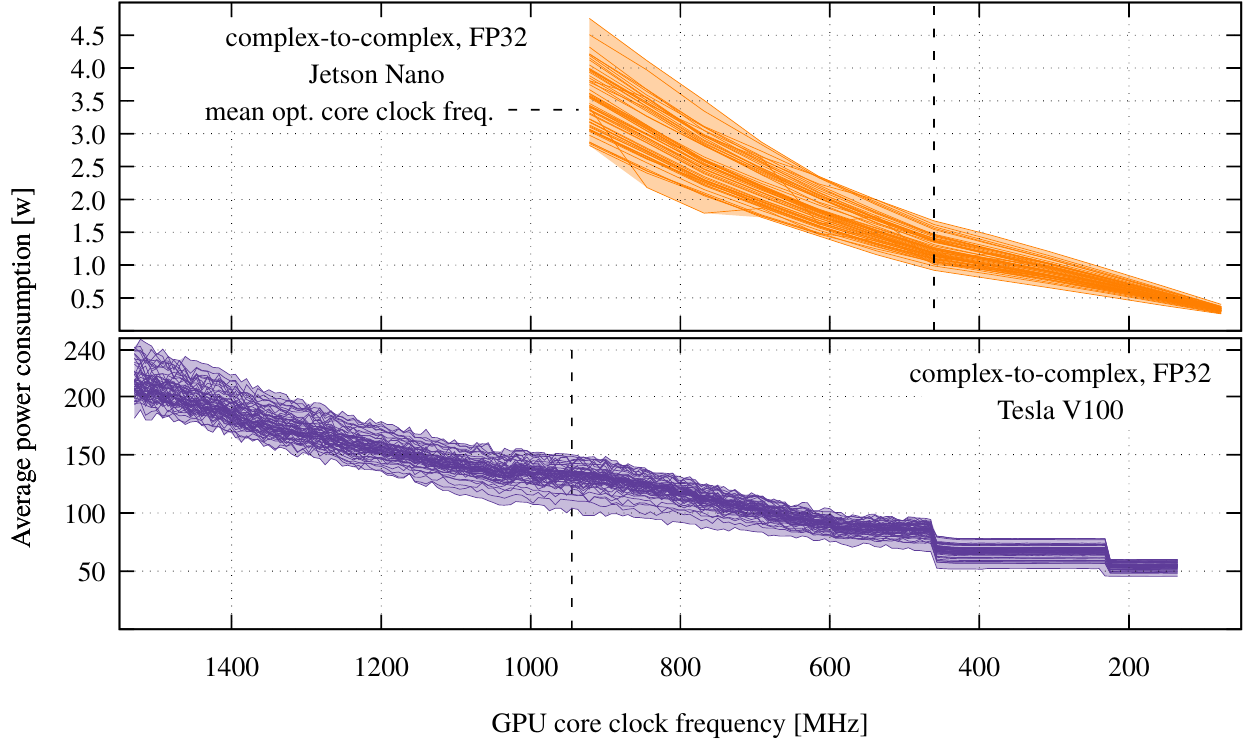}
    \caption{Averaged power consumption as a function of core clock frequency for all tested FFT lengths. The Jetson Nano is shown independently as its behaviour is different from the rest of the tested GPUs which are represented by the V100 GPU.\label{fig:wattfreq}}
\end{figure}

As the core clock frequency decreases the power consumption of 0the GPU changes non-linearly. This is shown in Fig.~\ref{fig:wattfreq} for the V100 GPU and Jetson Nano.

\begin{figure}[h!]
    \centering
    \includegraphics[width=\linewidth]{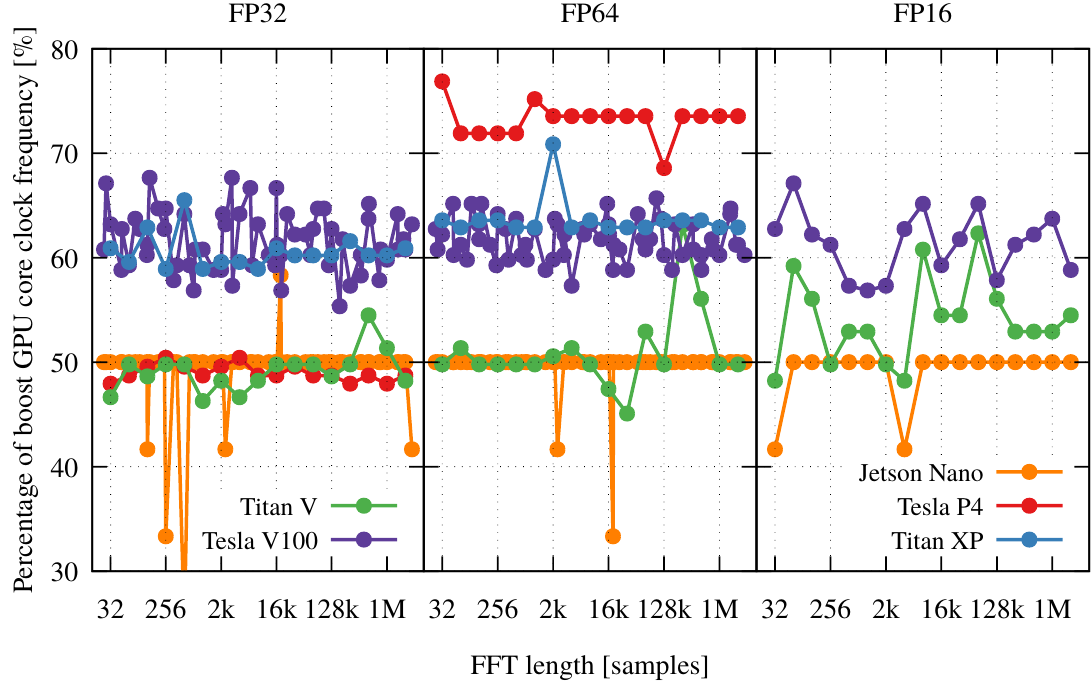}
    \caption{Value of the optimal frequency expressed as a percentage of the boost clock frequency. The value of the optimal frequency is consistent through different precisions with the exception of the Tesla P4 GPU. \label{fig:optimalfrequency}}
\end{figure}

The frequency at which the energy per FFT batch reaches a minimum was selected as the \textit{optimal frequency}. The optimal frequency is different for each tested FFT length for a given GPU and precision. The optimal frequency expressed as a percentage of the default core clock frequency for all precisions is shown in Fig.~\ref{fig:optimalfrequency}.
 
\subsection{Energy savings}
\begin{figure}[htp]
	    \begin{minipage}[t]{.45\textwidth}
            \centering
            \includegraphics[width=\linewidth]{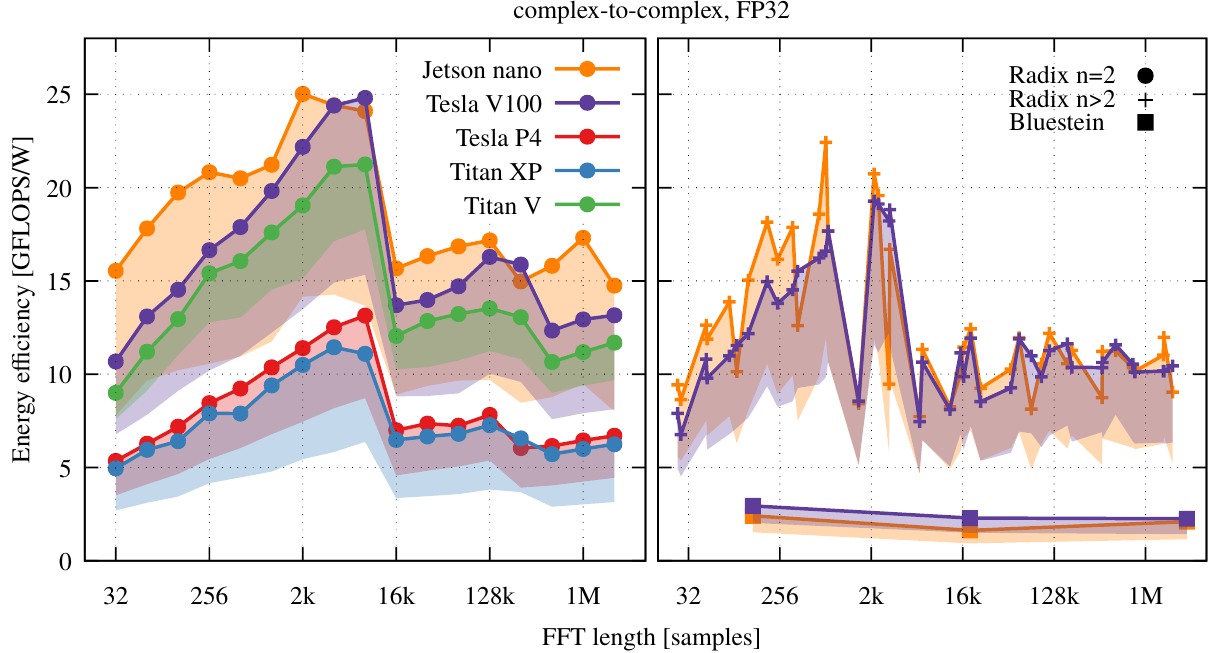}
	    \end{minipage}\\[0.2cm]
	    \begin{minipage}[t]{.45\textwidth}
		    \centering
    		\includegraphics[width=\linewidth]{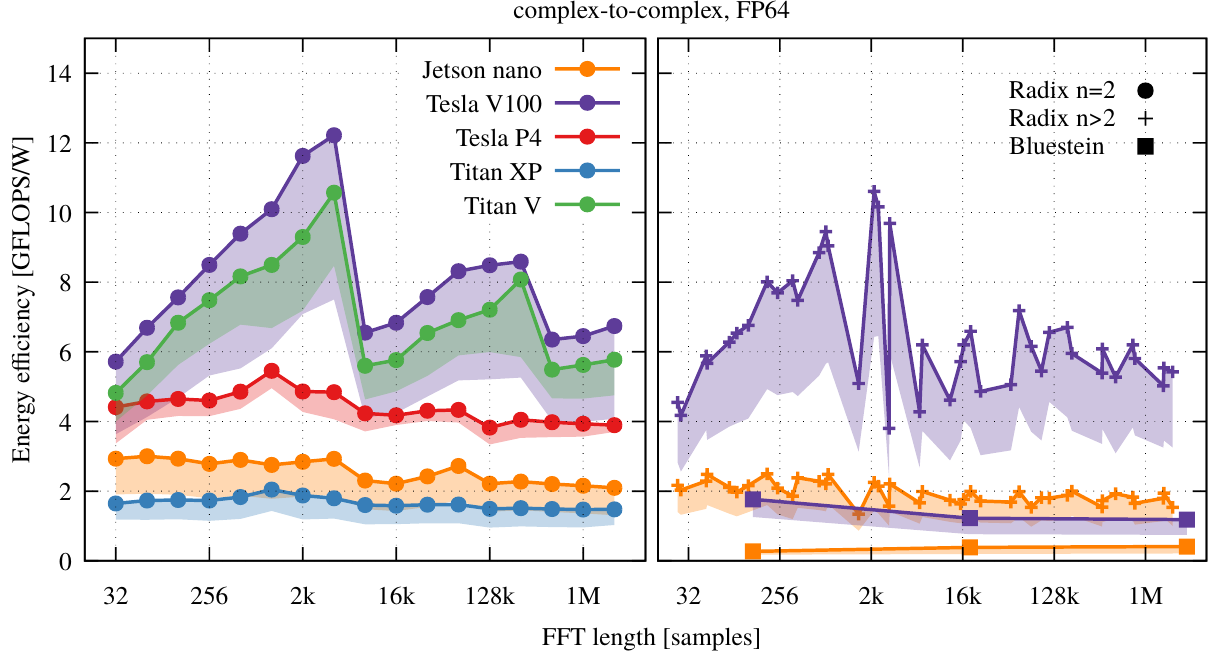}
	    \end{minipage}\\[0.2cm]
	    \begin{minipage}[t]{.45\textwidth}
            \centering
            \includegraphics[width=\linewidth]{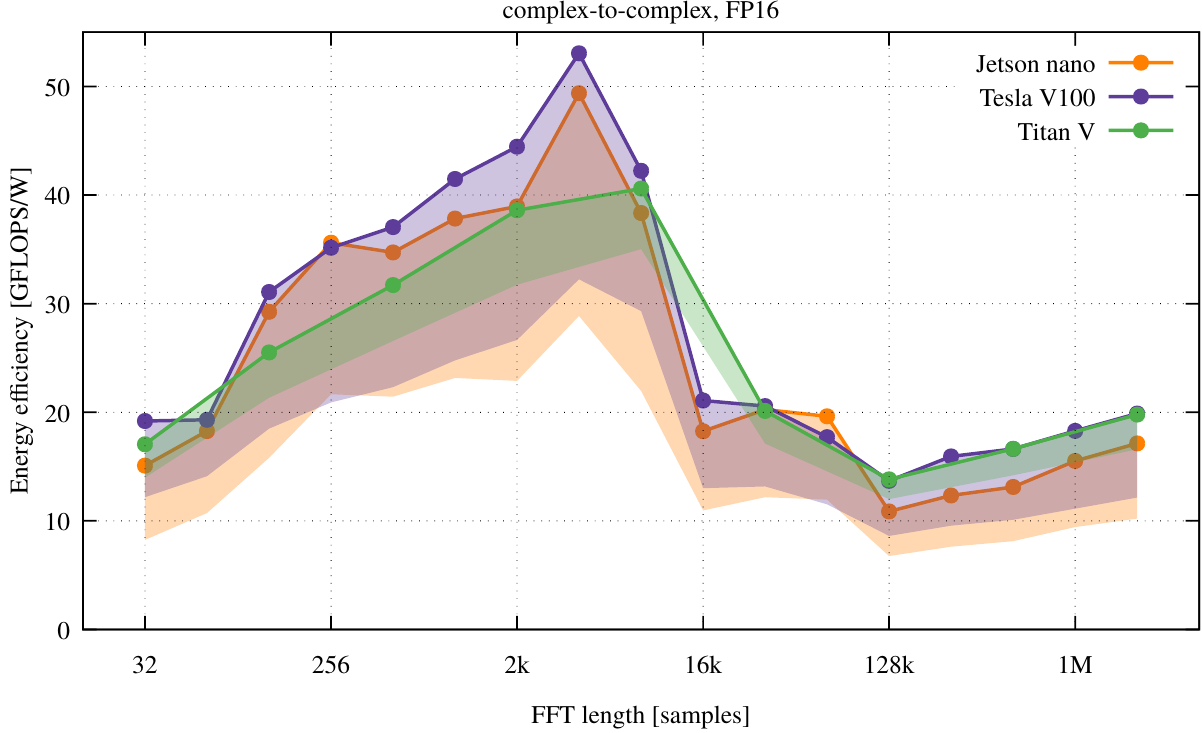}
	    \end{minipage}\\[0.2cm]
	    \caption{Floating-point operations per second per Watt (GFLOPS/W) for optimal frequency. The coloured region shows the improvement from the default frequency.}
	    \label{fig:gflop_per_watt}
\end{figure}

To acquire the following results we have selected the optimal frequency for each FFT length and measured the consumed power to calculate the energy efficiency using equation~\eqref{eq:efficiency}. The energy efficiency expressed as the number of GFLOPS/W is shown in Fig.~\ref{fig:gflop_per_watt}.

\begin{figure}[htp]
	    \begin{minipage}[t]{.45\textwidth}
            \centering
            \includegraphics[width=\linewidth]{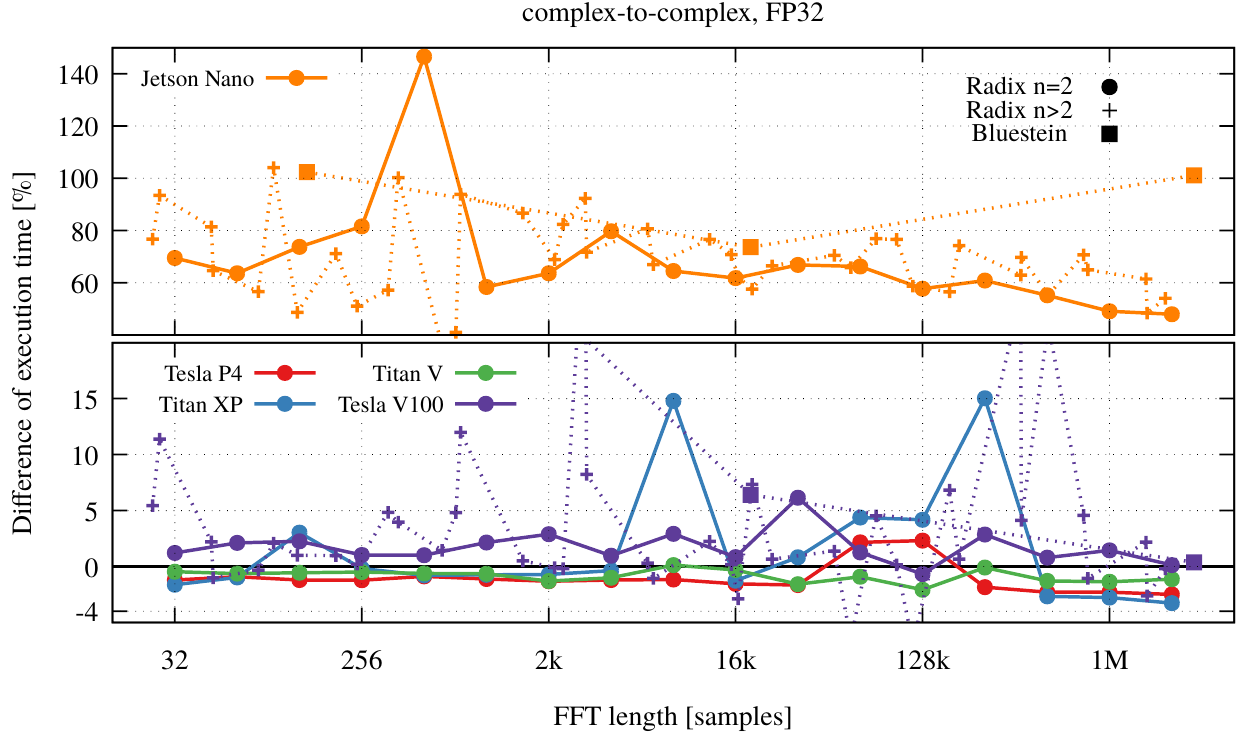}
	    \end{minipage}\\[0.2cm]
	    \begin{minipage}[t]{.45\textwidth}
		    \centering
    		\includegraphics[width=\linewidth]{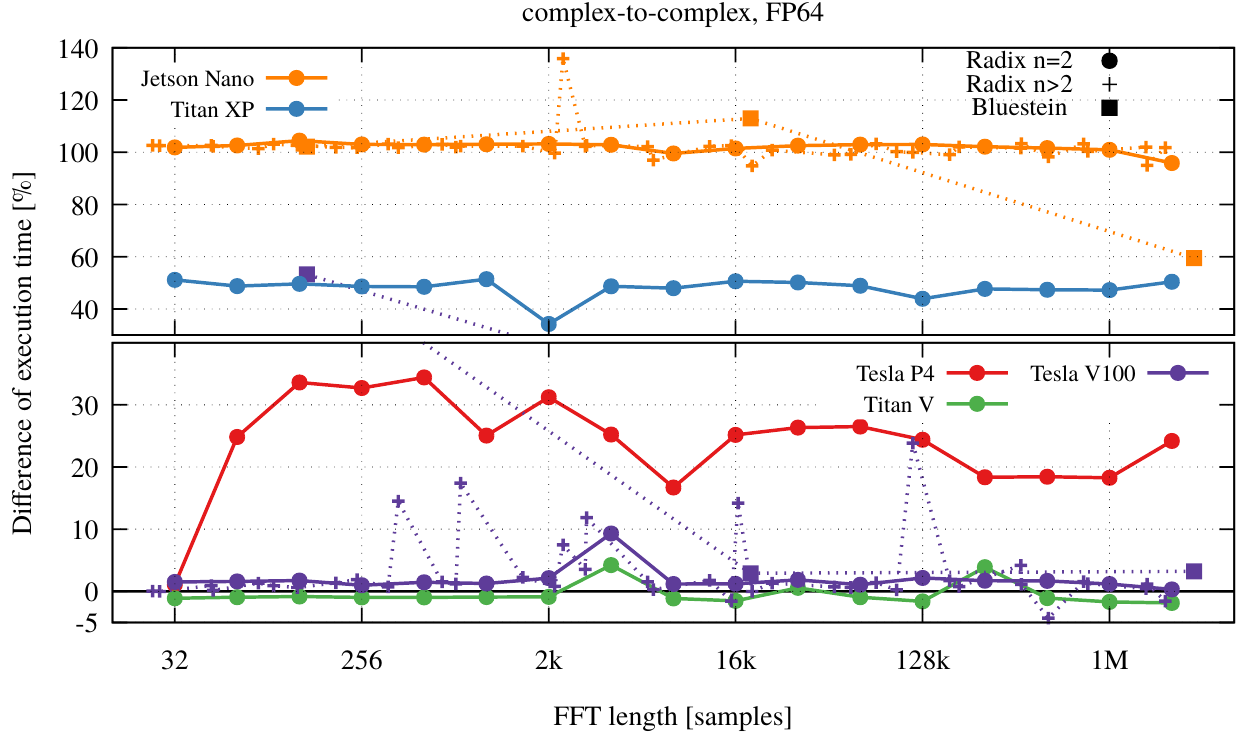}
	    \end{minipage}\\[0.2cm]
	    \begin{minipage}[t]{.45\textwidth}
            \centering
            \includegraphics[width=\linewidth]{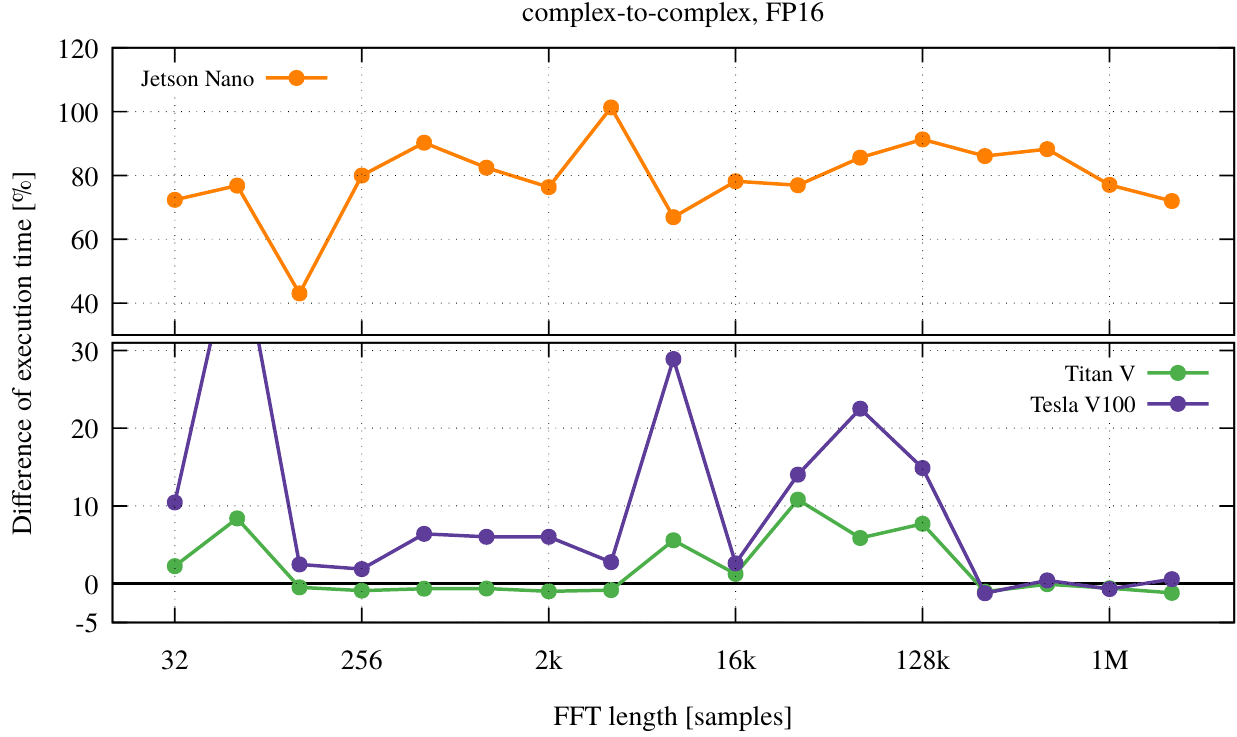}
	    \end{minipage}\\[0.2cm]
	    \caption{Increase in the execution time for optimal frequencies as a percentage of the default execution time $\td$.}
	    \label{fig:exectimediff}
\end{figure}

\begin{figure}[htp]
	\begin{minipage}[t]{\linewidth}
		\centering
		\includegraphics[width=\linewidth]{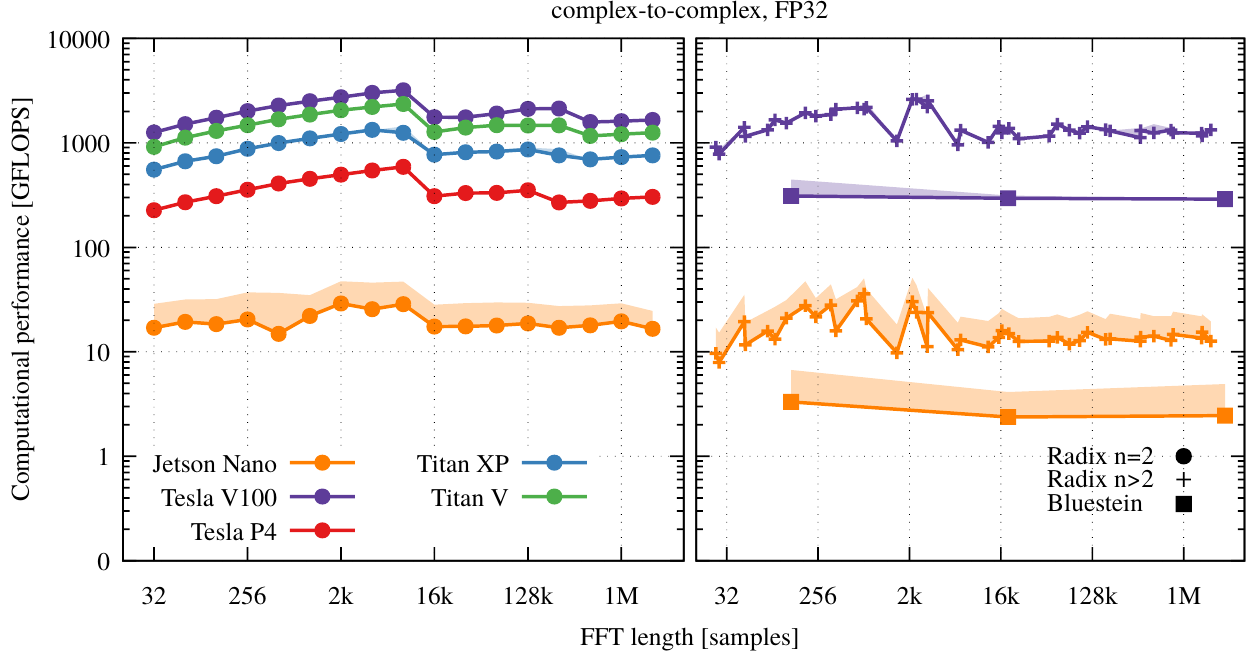}
	\end{minipage}\\[0.2cm]
	\begin{minipage}[t]{\linewidth}
		\centering
		\includegraphics[width=\linewidth]{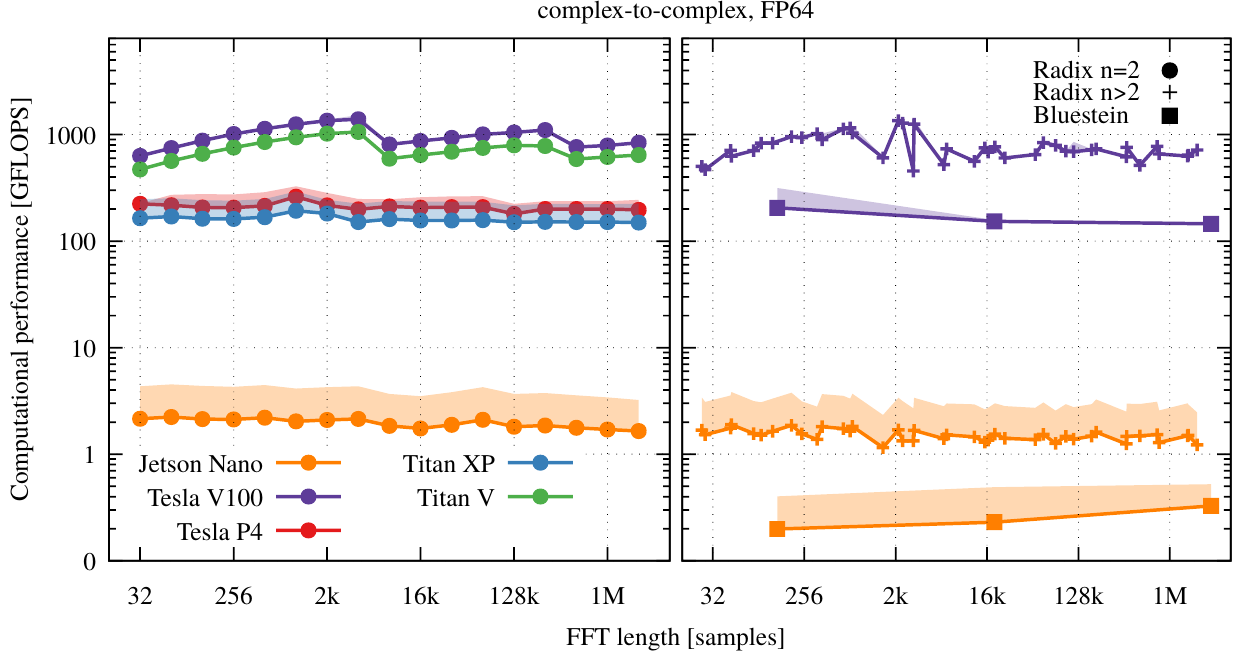}
	\end{minipage}\\[0.2cm]
	\begin{minipage}[t]{\linewidth}
		\centering
		\includegraphics[width=\linewidth]{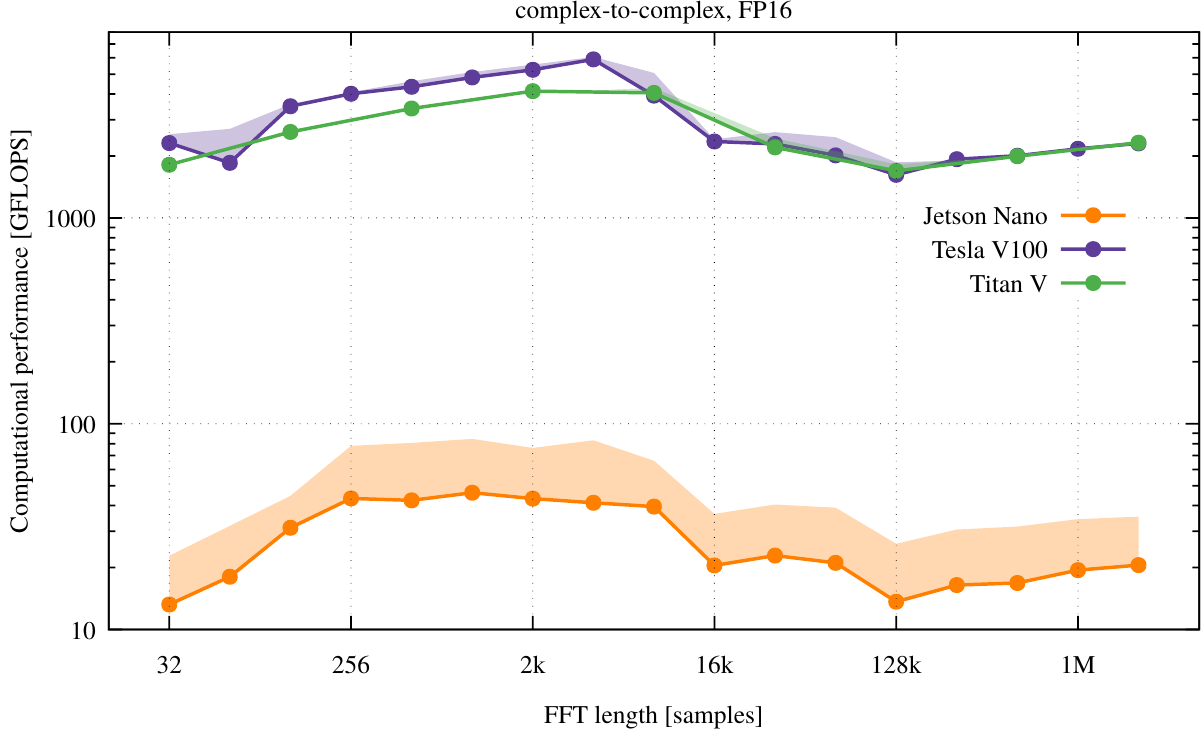}
	\end{minipage}\\[0.2cm]
	\caption{Floating-point operations per second (GFLOPS) for optimal frequencies. The colored region shows the change from the default frequency.}
	\label{fig:gflopchange}
\end{figure}

The change in the execution time for the optimal frequency with respect to the default execution time as a percentage is shown in Fig.~\ref{fig:exectimediff}. The change in GFLOPS is shown in Fig.~\ref{fig:gflopchange}. The peaks visible in Fig.~\ref{fig:exectimediff} correspond to FFT lengths which displayed case c) type behaviour of the execution time (Fig.~\ref{fig:timefreqdep}).

\begin{figure}[h!]
    \centering
    \includegraphics[width=\linewidth]{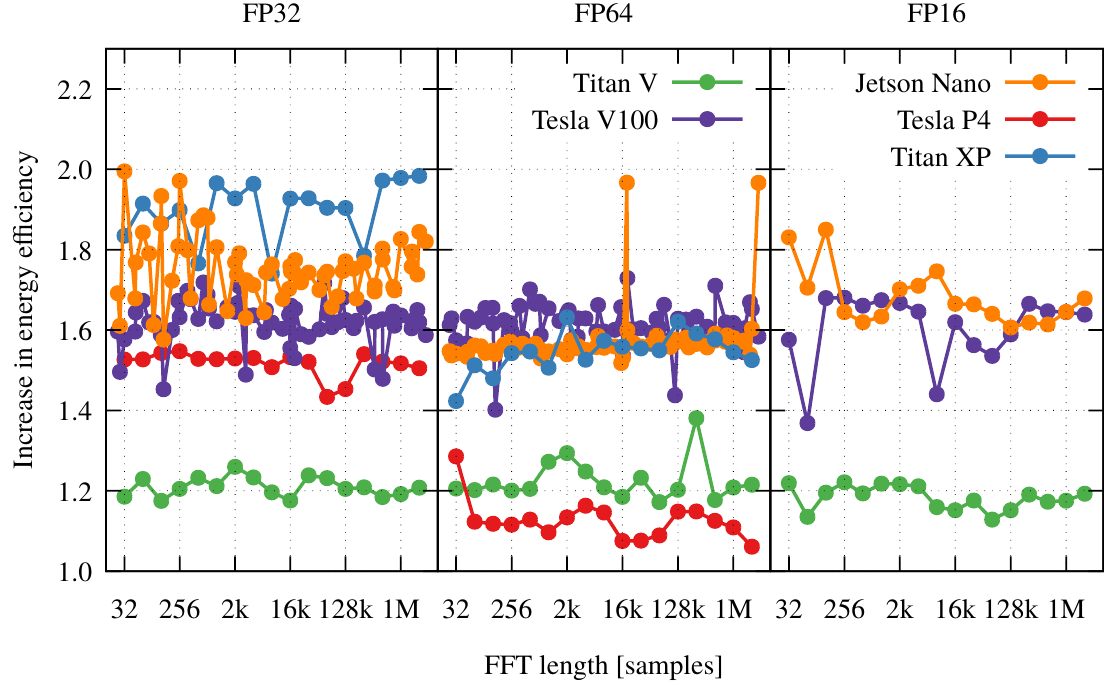}
    \caption{The increase in the energy efficiency for optimal core clock frequencies with respect to the \textbf{boost core clock frequency} for all tested FFT lengths. The two peaks observed in the Jetson Nano data are due to the use of the Bluestein algorithm.  \label{fig:energyeffincrease_boost}}
\end{figure}

\begin{figure}[h!]
    \centering
    \includegraphics[width=\linewidth]{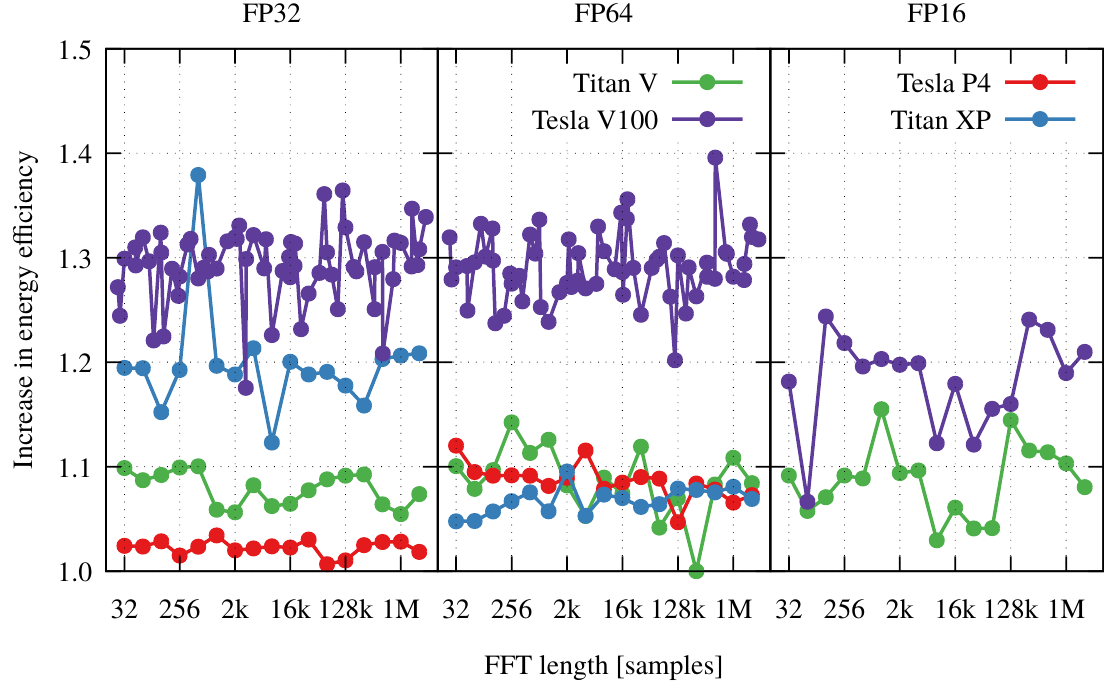}
    \caption{The increase in the energy efficiency for the optimal core clock frequencies with respect to the \textbf{base core clock frequency} for all tested FFT lengths. The Jetson Nano is not included since there is no base core clock frequency. \label{fig:energyeffincrease_base}}
\end{figure}

The increase in the energy efficiency \eqref{eq:increase} with respect to the boost core clock frequency is shown for different precisions in Fig.~\ref{fig:energyeffincrease_boost} and with respect to the base core clock frequency in Fig.~\ref{fig:energyeffincrease_base}.

\begin{figure}[h!]
    \centering
    \includegraphics[width=\linewidth]{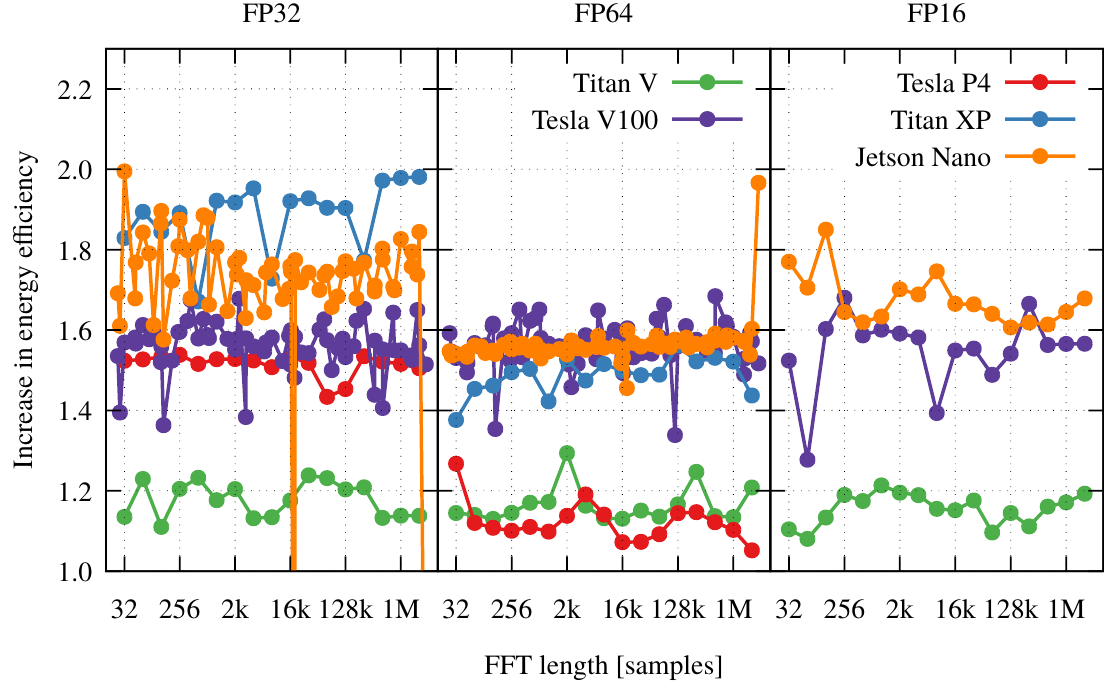}
    \caption{The increase in the energy efficiency for the mean optimal frequency with respect to the \textbf{boost core clock frequency} for all tested FFT lengths.The two peaks observed in the Jetson Nano data are due to the use of the Bluestein algorithm. \label{fig:mean_iee_boost}}
\end{figure}

\begin{figure}[h!]
    \centering
    \includegraphics[width=\linewidth]{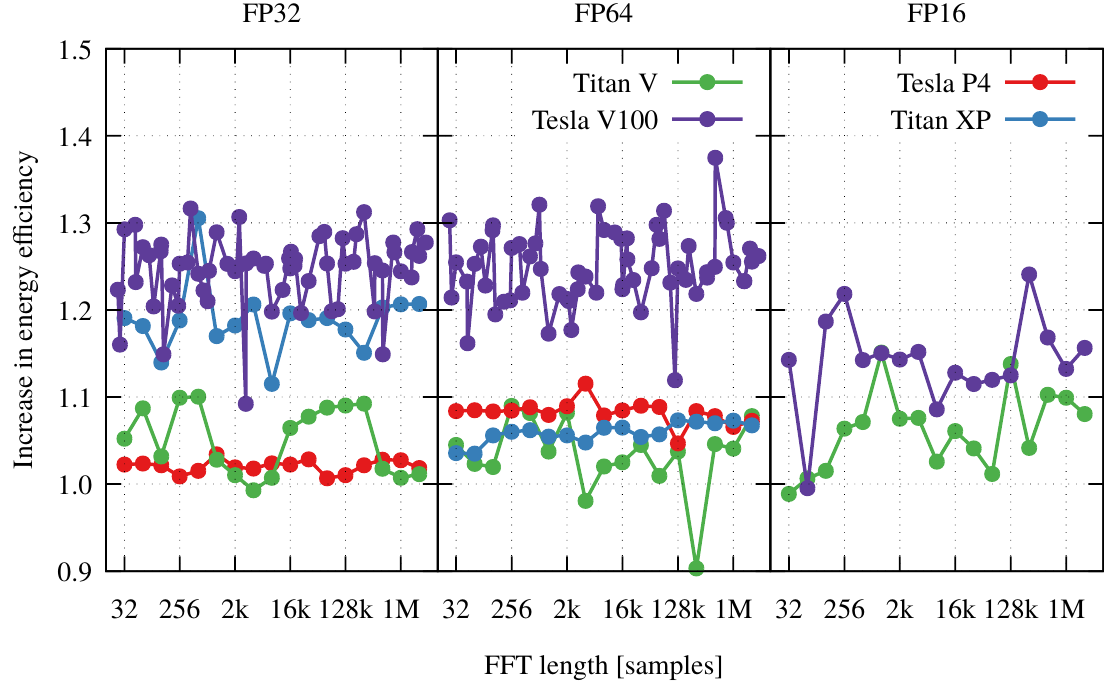}
    \caption{The increase in the energy efficiency for the mean optimal frequency with respect to the \textbf{base core clock frequency} for all tested FFT lengths. The Jetson Nano is not included since there is no base core clock frequency. \label{fig:mean_iee_base}}
\end{figure}

We see that the optimal frequency of different FFT lengths as shown in Fig.~\ref{fig:optimalfrequency} is roughly the same for a given GPU and precision across all tested FFT lengths. Furthermore, the optimal frequency is roughly the same across all numerical precisions for a given GPU with the exception of Tesla P4 GPU. Based on this we have calculated a \textit{mean optimal frequency} for a given GPU and precision by averaging optimal frequencies which achieves a similar increases in energy efficiency for all measured FFT lengths. The increase in energy efficiency using the mean optimal frequency is shown in Fig.~\ref{fig:mean_iee_boost} for the boost frequency and in Fig.~\ref{fig:mean_iee_base} for the base frequency. The values of mean optimal frequencies are listed in Table~\ref{tab:meanoptimal}.

\begin{table}[h]
\centering
\caption{\label{tab:meanoptimal}Mean optimal core clock frequencies.}
\resizebox{\linewidth}{!}{
\begin{tabular}{|l|r|r|r|}
\hline
Card name   & FP32 [MHz]  & FP64 [MHz] & FP16 [MHz]  \\
\hline
Tesla V100  & 945   & 945   & 937   \\
Tesla P4    & 746   & 1126  & NA    \\
Titan V     & 952   & 967   & 1042  \\
Titan XP    & 1151  & 1215  & NA    \\
Jetson Nano & 460.8 & 460.8 & 460.8 \\
\hline
\end{tabular}
}
\end{table}

\begin{figure}[h!]
    \centering
    \includegraphics[width=\linewidth]{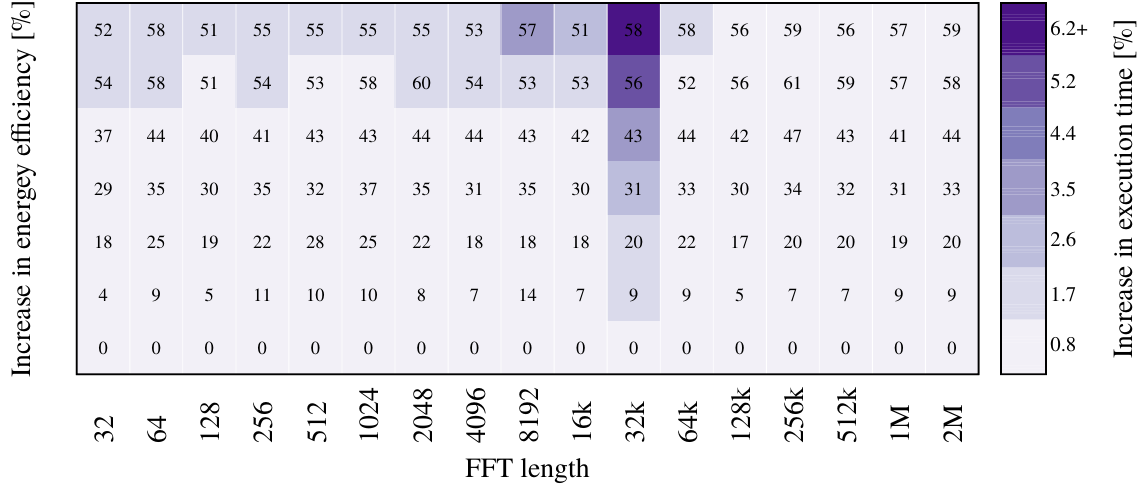}
    \caption{Trade-off between an increase in energy efficiency in percent (represented by a number in each cell) and an increase in execution time (represented by a color) for the V100 GPU. \label{fig:iee_et_V100}}
\end{figure}

\begin{figure}[h!]
    \centering
    \includegraphics[width=\linewidth]{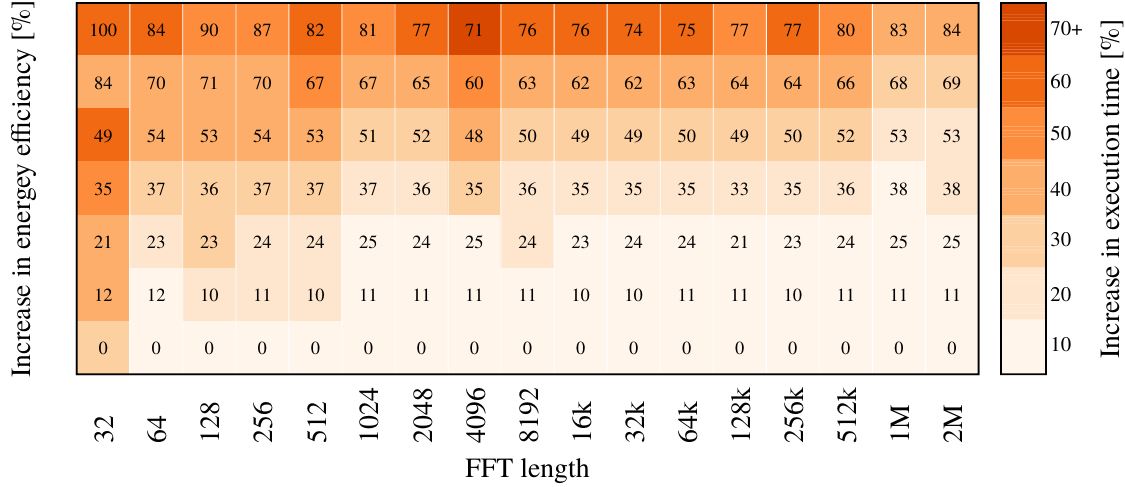}
    \caption{Trade-off between an increase in energy efficiency in percent (represented by a number in each cell) and an increase in execution time (represented by a color) for the Jetson Nano. \label{fig:iee_et_jetson}}
\end{figure}

When considering existing pipelines, it is also interesting to study the relationship between the increase in energy efficiency and the increase in the execution time. This relationship indicates the cost (in units of execution time) of any increase in energy efficiency. This is shown for the V100 GPU in Fig.~\ref{fig:iee_et_V100} and for the Jetson Nano in Fig.~\ref{fig:iee_et_jetson}.

\subsection{Integration into existing pipe\-lines}
To demonstrate the applicability of the mean optimal frequency in existing pipelines we have employed part of the data processing pipeline\footnote{Source code for used pipeline is on GitHub \url{https://github.com/KAdamek/cuFFT_energy_efficiency_example}} used for the detection of pulsars in time-domain radio astronomy data. The pipeline uses several computational steps: FFT, power spectrum calculation; mean and standard deviation calculation; and the harmonic sum. The harmonic sum adds the value of higher harmonics of the pulsar in the power spectrum to the pulsar's expected fundamental frequency thus increasing the signal-to-noise ratio of the pulsar in the power spectrum. The harmonic sum can add up to 32 higher order harmonics which decreases the FFT execution footprint in the pipeline's total execution time.

\begin{figure}
	\centering
	\includegraphics[width=\linewidth]{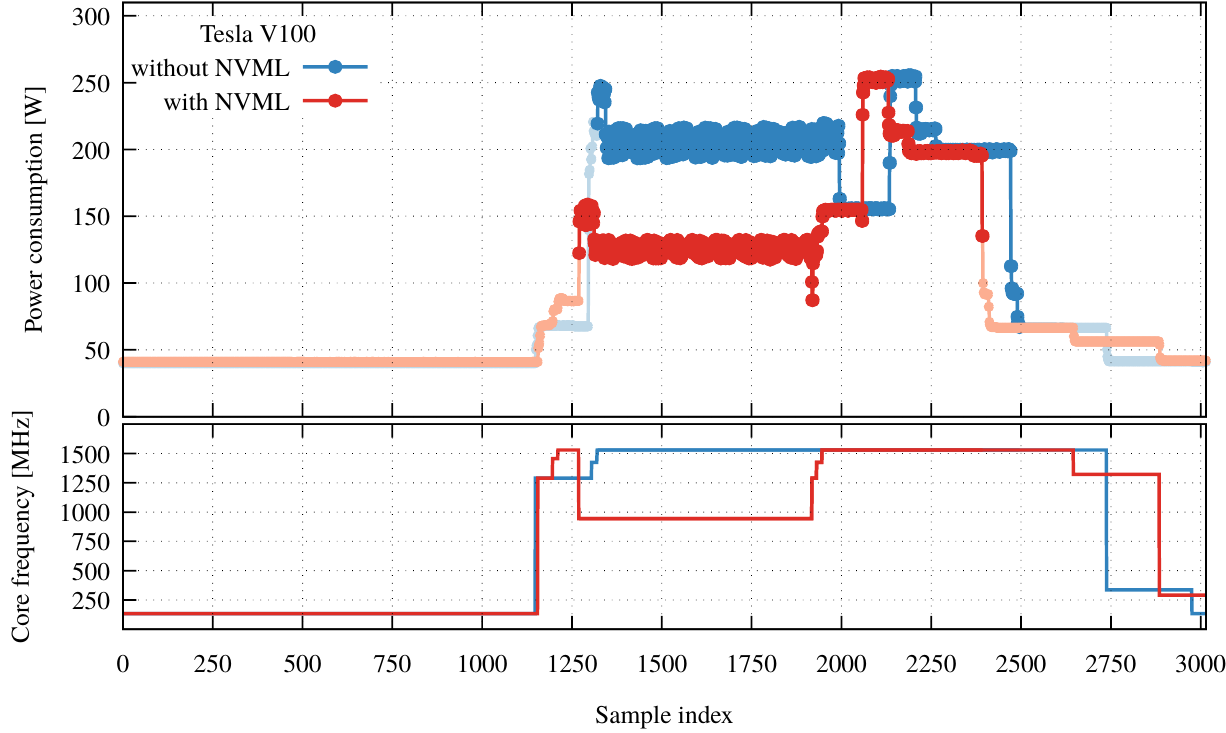}
	\caption{Measured power consumption (top) and core clock frequency (bottom) for part of a radio astronomy data processing pipeline. \label{fig:pipeline_harmonics}}
\end{figure}

To change the frequency during the pipeline execution we have used the NVIDIA Management Library (NVML) \cite{web:NVIDIA_NVML}. This approach, however, has limitations because the library is fully supported only on scientific (Tesla) NVIDIA GPUs. The measured power consumption and the core clock frequency for the V100 GPU are shown in Fig. \ref{fig:pipeline_harmonics} and the increase in energy efficiency for different configurations of the pipeline is listed in Table \ref{tab:pipeline_harmonics}. 

The usage of the NVML library is simple. Before the GPU kernel execution the core clock frequency is (for a given GPU) set using \texttt{nvmlDeviceSetGpuLockedClocks} providing maximum and minimum core clock frequency. When the calculation is finished the GPU core clock frequency is returned to default by calling \texttt{nvmlDeviceResetGpuLockedClocks}.

The FFT length used for the computation was $N=5\cdot10^5$ which was not used in our measurements or in our calculation of the mean optimal frequency.

\begin{table}[ht]
\centering
\caption{Increase in energy efficiency for different configurations of our toy data processing pipeline. \label{tab:pipeline_harmonics}}
\begin{tabular}{ccc}
\toprule
\multicolumn{1}{p{.2\linewidth}}{num. harmonic summed}& \multicolumn{1}{p{.25\linewidth}}{cuFFT \% of total exec. time} & \multicolumn{1}{p{.3\linewidth}}{Increase in Energy efficiency} \\
\midrule
2                    & 60.85                                      & 1.291                         \\
4                    & 58.56                                      & 1.290                         \\
8                    & 55.92                                      & 1.267                         \\
16                   & 53.73                                      & 1.260                         \\
32                   & 51.34                                      & 1.240                        \\
\bottomrule
\end{tabular}
\end{table}

\subsection{Profiling}
\begin{figure*}
	\centering
	\includegraphics[width=0.85\textwidth]{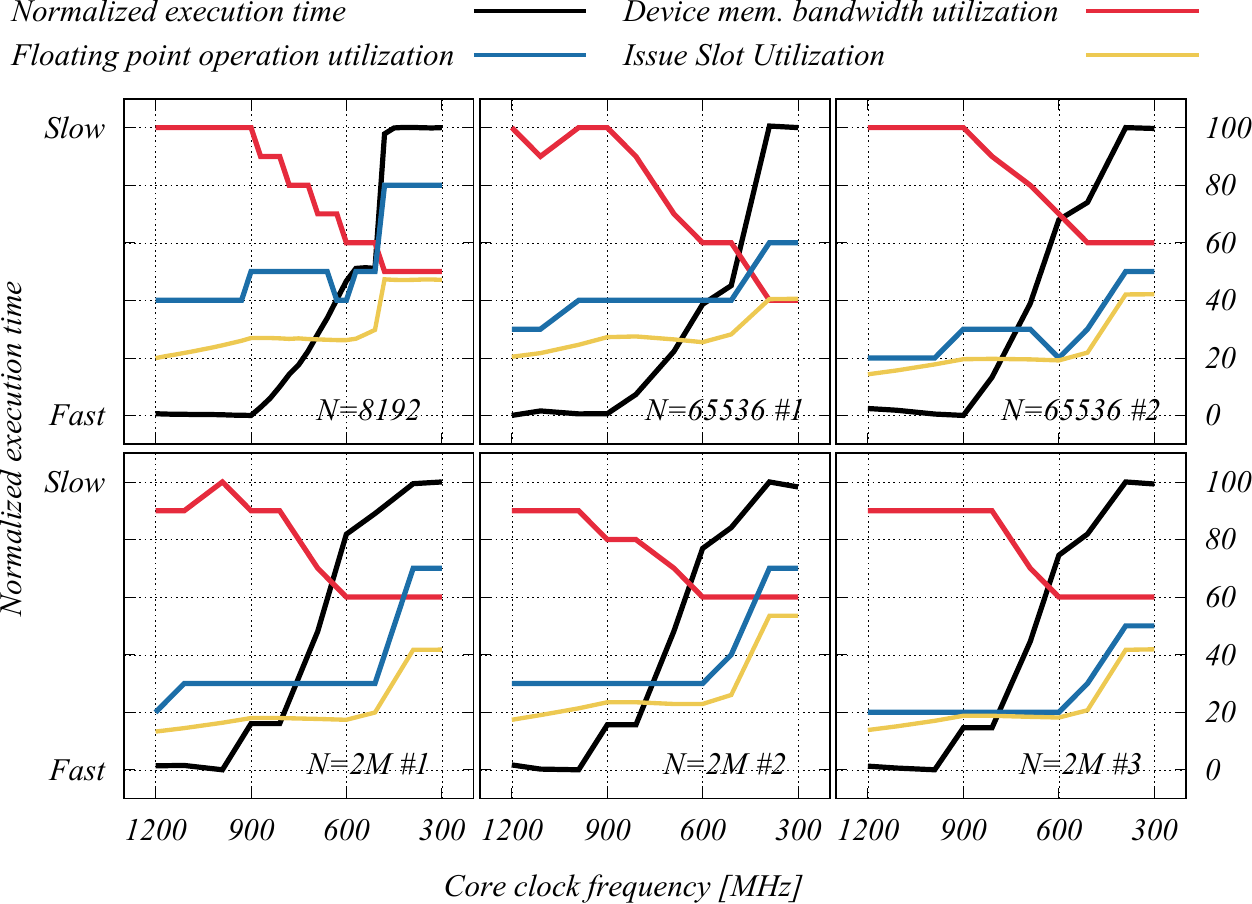}
	\caption{Profiling results for the V100 GPU using the NVIDIA visual profiler. Longer FFT lengths use more than one GPU kernel to calculate the Fourier transform which are numbered. \label{fig:profiling}}
\end{figure*}

For profiling, we have used the NVIDIA visual profiler (NVVP). Based on the different behavior of the execution time $\tfix$ shown in Fig.~\ref{fig:fixedmemExTFP32} we have selected three representative power-of-two FFT lengths ($N=8192$, $N=16\mathrm{k}$, $N=2\mathrm{M}$) which are calculated by different kernels. The profiling results for these kernels are shown in Fig.~\ref{fig:profiling}. For our study of compute utilization we have used two indicators. The first is the compute utilization as reported by the NVVP, the second metric is the issue slot utilization, which tells us how many instruction slots are used. The next quantity displayed in Fig.~\ref{fig:profiling} is the device memory bandwidth utilization (device MBU). Fig.~\ref{fig:profiling} also shows the normalized execution time from fastest to slowest to provide context for the other displayed quantities.

\section{Discussion}
\label{sec:discussion}
The dependency of the execution time on the core clock frequency is shown in Fig.~\ref{fig:timefreqdep}. Fig.~\ref{fig:timefreqdep} displays the three previously discussed behaviours a), b) and c). However, the Jetson Nano only exhibits the third type of behaviour c). All other GPU's, represented by the V100 GPU, exhibit a composition of all three behaviours with cases a) and b) being dominant.

The behaviour in case b), might be due to reduced cache contention which slightly increases the hit rate of the unified cache as shown by the the NVVP. However, it might also be a systematic error caused by measurement using the NVIDIA driver, which is based on the GPU core clock frequency. In this case as well as in case a) the GPU's compute resources are not fully utilized and the computations are limited by device memory bandwidth. 

The reason for an increase in the execution time at a particular critical frequency is due to the saturation of the number of issued instructions (see Fig.~\ref{fig:profiling}). This leads to a reduction in memory requests to the device memory which, in turn, leads to poor latency hiding of the device memory accesses. Therefore most of the threads are waiting for data but there are not enough threads with data to utilize the floating-point operation units. Thus the floating-point operation utilization remains mostly unchanged.

The sharp increase in the execution time $\tfix$ for low frequencies, which are present in all cases, are due to the change of the P-state to a state corresponding to the idle status of the GPU with reduced voltage which reduces the available GPU resources severely.  

Lastly, case c) occurs due to the high utilization of one of the caches. Since the cache bandwidth decreases with the core clock frequency each decrease in frequency lowers the bandwidth which is already fully utilised leading to a decrease in performance.

The average power consumption shown in Fig.~\ref{fig:wattfreq}, tells us why, even with longer execution times, we can improve energy efficiency. The rate of the decrease in power consumption is higher than the rate at which the execution time increases. This is especially visible around $f=1000\,\mathrm{Hz}$ for the V100 GPU and about $f=450~\mathrm{Hz}$ for the Jetson Nano. These frequencies roughly coincide with the mean optimal frequency for the given GPUs.

\subsection{Real-time processing}
The energy efficiency is shown in Fig.~\ref{fig:gflop_per_watt}, the change in the execution time is shown in Fig.~\ref{fig:exectimediff} and the change in GFLOPS is shown in Fig.~\ref{fig:gflopchange}. 

In the language of costs, Fig.~\ref{fig:exectimediff} is equivalent to the increase in capital costs as an increase in execution time directly translates into more hardware needed in order to meet the constrains of real-time data processing. On the other hand, the increase in energy efficiency (Fig.~\ref{fig:gflop_per_watt}) is related to operational costs, where better energy efficiency translates into lower operational costs. However we must bare in mind that operational costs include cooling, facility management, etc. which could be increased by the requirement for more hardware due to longer execution times.

For FP32 precision we see that the Jetson Nano is more energy efficient than the V100 GPU for almost all FFT lengths, especially for the small FFT lengths where it is 50\% more efficient. When we look at the change in the execution time we see that the Jetson Nano requires approximately 60\% more time to finish compared to the execution time at the boost core clock frequency. With one extreme case where the execution time is 140\% longer. This means on average 60\% more hardware to achieve real-time data processing with the best energy efficiency.

This behaviour is not reproduced by the V100 GPU where the increase in energy efficiency is not, for the most part, at the expense of the execution time. The change in the execution time for the V100 GPU is below 5\%. There are more significant increases in execution time for the non-power of two FFT lengths which can cause increases of up to 20\% in execution time. Small changes in the execution time on the V100 GPU offers a possibility to improve existing real-time processing pipelines without substantial change in hardware.

We see similar behaviour for the V100 GPU at FP64 precision. The slow-down in execution time suffered by the V100 GPU due to the lower core clock frequencies is within 5\%. The execution time for most of the non-power of two FFT lengths does not increase above 20\%. The Tesla P4 GPU, Titan XP GPU and Jetson Nano do not fully support FP64 precision. This manifests in less significant improvements in GFLOPS/W, much higher execution times and a decrease in GFLOPS. In the case of the Jetson Nano we would have to double the number of cards in order to process data in real-time.

At FP16 precision we have only three GPUs which support this precision: V100 GPU, Titan V GPU and Jetson Nano. Regarding energy efficiency, both the V100 GPU and the Jetson Nano are comparable but the V100 GPU is the overall more energy efficient GPU. When we look at the change in execution time we see that the V100 GPU typically has a 10\% increase or less, but at some FFT lengths the increase is as high as 40\% (N=64). This behaviour means that we have to be more careful about potential energy savings since at some FFT lengths the increase in execution time might be too high for real-time data processing. The change in execution time of the Jetson Nano is again large and we would need to have almost twice the number of GPUs to process data in real time at the best possible energy efficiency.

\subsection{Increase in energy efficiency}
The increase in the energy efficiency for the optimal frequency is shown in Fig.~\ref{fig:energyeffincrease_boost} and Fig.~\ref{fig:energyeffincrease_base}. The corresponding figures for the mean optimal frequency are Fig.~\ref{fig:mean_iee_boost} and Fig.~\ref{fig:mean_iee_base}. The difference in the increase in energy efficiency for the base core clock frequency between the optimal frequency and the mean optimal frequency is 5 percentage points. That is, an average increase in energy efficiency for the optimal frequency which is tuned for each FFT length is 29\% whereas the average increase in energy efficiency for the mean optimal frequency is 24\%. For the V100 GPU this holds for all FFT lengths and precisions with a very limited number of exceptions for FP16 precision. For the boost core clock frequency the loss is 10 percentage points. This allows us to use one core clock frequency and achieve similar energy savings without determining the optimal frequency for each FFT length. A similar result is observed for the Jetson Nano with the exception of Bluestein FFT lengths which are responsible for the peaks in the results.

The dependency between the increase in energy efficiency and the change in the execution time, shown in Fig.~\ref{fig:iee_et_V100} for the V100 GPU but more notably in Fig.~\ref{fig:iee_et_jetson} for the Jetson Nano, is non-linear. We see that we can achieve an interesting increase in energy efficiency even for increases in execution time which are below 10\%. 

Lastly, our practical test with our example data processing pipeline shows that we can dynamically change the core clock frequency in a very precise manner. Our code demonstrates how to target only the duration of the cuFFT library call within the pipeline and thus reduce power consumption. This technique can be applied to existing pipelines or more generally any software with minimal changes to the codebase. The increase in energy efficiency (for the boost core clock frequency) are summarized in Table~\ref{tab:pipeline_harmonics} corresponds to the expected values based on the FFT execution time footprint within the pipeline. For the first configuration with 2 harmonics, the FFT execution time corresponds to 60\% of the total execution time. The average increase in energy efficiency for V100 GPU with boost core clock frequency (based on Fig.\ref{fig:mean_iee_boost}) is about 50\%. Considering the FFT execution time footprint we should get 30\% increase in energy efficiency which is indeed what we have measured. This behaviour is consistent with other configurations of the pipeline.

\section{Conclusions}
\label{sec:conclusions}
We have measured the power consumption when calculating the Fourier transformation at different numerical precisions (FP32, FP64, FP16) on NVIDIA GPUs using the NVIDIA cuFFT library and quantified the possible energy savings when DVFS techniques are used. For each tested GPU, precision, and a wide range of FFT lengths, we have found the optimal core clock frequency to minimise power consumption. We have also measured the change in execution time of the Fourier transform when DVFS is applied, which is an important consideration for real-time data processing because this can increase when the core clock frequencies of the GPU are modified. 

We have presented the achieved energy efficiency in GFLOPS/W. Along with this we have presented the increase in energy efficiency when using our optimal core clock frequency compared to the boost and base core clock frequency for each GPU. We have also presented the increase in the execution time of the Fourier transform when DVFS is applied.

The decrease in power consumption and change in the execution time depends on the GPU used. In the case of the V100 GPU, the average increase in energy efficiency is for FP32, FP64, and FP16 precisions is 60\% compared to the boost core clock frequency. When compared to the base core clock frequency an average increase in energy efficiency of 30\% for FP32 and FP64 precision and 20\% for FP16 precision is observed. The increase in the execution time is below 5\% (with few exceptions as outlined). The Jetson Nano offers higher increases in energy efficiency to that of the V100 GPU. On average 70\% for FP32, 55\% for FP64 and 70\% for FP16 but at the expense of execution time which increases by more than 60\%. For the P4 GPU and the Titan V GPU we have not achieved a significant increase in energy efficiency.

Our results have shown that the Volta architecture is significantly more energy efficient than the P4 GPU which represents the most energy efficient GPU from the previous Pascal generation. When compared to the Jetson Nano the V100 GPU is less energy efficient at FP32 precision. For short and long FFTs at FP32 precision the Jetson Nano is 50\% more energy efficient than the V100 GPU. For FP16 precision the V100 GPU has similar energy efficiency as the Jetson Nano. The Jetson Nano does not fully support double precision thus the V100 GPU is significantly more energy efficient at this precision. 

We have shown that values of optimal core clock frequencies for all tested FFT lengths for a given GPU and numerical precision are similar, with few exceptions. This allowed us to define a mean optimal core clock frequency unique to each tested GPU and precision, but is the same for all FFT lengths. Using the mean optimal core clock frequency, we have achieved a similar energy efficiency when compared to the energy efficiency achieved with the optimal core clock frequency for each tested FFT length. For the V100 GPU the difference is only 5 percentage points. For the other GPUs the loss is similar.

We have also presented the practical implementation of these results in our example data processing pipeline which is available as an open source code. We have demonstrated how to change the core clock frequency of the GPU to the mean optimal core clock frequency using the  NVIDIA Management Library and demonstrated a decrease in power consumption which is in agreement with the results presented in this work.

Finally we have highlighted how, from an environmental perspective, increasing the energy efficiency of the FFT algorithm will be an important consideration for edge computing and IoT. 

\section*{Acknowledgment}
This work has received support from STFC Grant (ST/T000570/1). The authors acknowledge the support of the OP VVV MEYS funded project CZ.02.1.01/0.0/0.0/16\_019/0000765 "Research Center for Informatics". The authors would like to acknowledge the use of the University of Oxford Advanced Research Computing (ARC) facility in carrying out this work (\url{http://dx.doi.org/10.5281/zenodo.22558}).
The authors would like to express their gratitude to the Research Centre for Theoretical Physics and Astrophysics, Institute of Physics, Silesian University in Opava for institutional support.

\bibliography{FFTeff}

\end{document}